%% file: main.tex
\documentclass[acmsmall, nonacm]{acmart}
\usepackage{multirow}
\usepackage{multicol}
\usepackage{arydshln}
\usepackage{multirow}
\usepackage{longtable}
\usepackage{array}
\usepackage{colortbl}
\usepackage{xcolor}
\usepackage{soul}
\usepackage{changepage}
\usepackage{enumitem}
\usepackage{arydshln}
\usepackage{graphicx}

\usepackage{booktabs}
\usepackage{tabularx}
\usepackage{ragged2e}
\newcolumntype{Y}{>{\RaggedRight\arraybackslash}X}
\usepackage[outdir=./]{epstopdf}

\definecolor{VA}{HTML}{ffebd2}
\definecolor{user}{HTML}{FFFFFF}
\definecolor{design}{HTML}{F7F8A7}
\definecolor{purple}{HTML}{d9cae8}

\sethlcolor{design} 
\definecolor{mediumgray}{HTML}{808080}

\newcolumntype{M}{>{\ttfamily\small}p{7.5cm}}
\newcolumntype{S}{>{\ttfamily\small}p{5.5cm}}
\newcolumntype{L}{>{\columncolor{VA}\ttfamily\small}p{7.5cm}}
\newcolumntype{X}{>{\columncolor{VA}\ttfamily\small}p{9.5cm}}
\newcolumntype{D}{>{\columncolor{design}\ttfamily\small}p{6cm}}
\newcolumntype{I}{>{\ttfamily\small}p{9.5cm}}

\newcommand{\myc}[2][mediumgray]{%
  \textcolor{#1}{#2}%
}

\def\eg{\textit{e.g., }}

\AtBeginDocument{%
  }

\setcopyright{acmlicensed}
\copyrightyear{2018}
\acmYear{2018}
\acmDOI{XXXXXXX.XXXXXXX}

\acmConference[Conference acronym 'XX]{Make sure to enter the correct
  conference title from your rights confirmation emai}{June 03--05,
  2018}{Woodstock, NY}
\acmISBN{978-1-4503-XXXX-X/18/06}

\begin{document}

\title{From Text to Conversation: Evaluating an LLM-Powered Voice Assistant for Sleep Diary Intake}

\title{Conversational Voice Sleep Diaries Improve Adherence and Contextual Richness in Longitudinal Self-Report}

\title{Conversational Voice Sleep Diaries Improve Adherence and Contextual Self-Report: A Four-Week Field Study}

\title{Better Adherence, Richer Context: A Field Evaluation of LLM-Powered Conversational Voice Diaries for Sleep}

\author{Amama Mahmood}
\authornote{Both authors contributed equally to this research.}
\email{amama.mahmood@jhu.edu}

\affiliation{%
  \institution{The Johns Hopkins University}
  \streetaddress{3400 N. Charles St}
  \city{Baltimore}
  \state{Maryland}
  \country{USA}
  \postcode{21218}
}

\author{Bokyung Kim}
\authornotemark[1]
\email{bkim94@jhu.edu}
\affiliation{%
  \institution{The Johns Hopkins University}
  \streetaddress{3400 N. Charles St}
  \city{Baltimore}
  \state{Maryland}
  \country{USA}
  \postcode{21218}
}

\author{Honghao Zhao}
\email{hzhao78@jhu.edu}
\affiliation{%
  \institution{The Johns Hopkins University}
  \streetaddress{3400 N. Charles St}
  \city{Baltimore}
  \state{Maryland}
  \country{USA}
  \postcode{21218}
}

\author{Molly E. Atwood}
\email{matwood4@jhmi.edu}
\affiliation{%
  \institution{Department of Psychiatry and Behavioral Sciences, The Johns Hopkins University School of Medicine}
  \city{Baltimore}
  \state{Maryland}
  \country{USA}
}

\author{Luis F. Buenaver}
\email{lbuenav1@jhmi.edu}
\affiliation{%
  \institution{Department of Psychiatry and Behavioral Sciences, The Johns Hopkins University School of Medicine}
  \city{Baltimore}
  \state{Maryland}
  \country{USA}
}

\author{Michael T. Smith}
\email{msmith62@jhmi.edu}
\affiliation{%
  \institution{Department of Psychiatry and Behavioral Sciences, The Johns Hopkins University School of Medicine}
  \city{Baltimore}
  \state{Maryland}
  \country{USA}
}

\author{Chien-Ming Huang}
\email{chienming.huang@jhu.edu}
\affiliation{%
  \institution{The Johns Hopkins University}
  \streetaddress{3400 N. Charles St}
  \city{Baltimore}
  \state{Maryland}
  \country{USA}
  \postcode{21218}
}

\renewcommand{\shortauthors}{Mahmood et al.}

\begin{abstract}
Sleep diaries are central to behavioral sleep medicine and cognitive behavioral therapy for insomnia, yet daily completion is difficult to sustain, and static forms often provide limited context for interpreting night-to-night sleep variation. We designed an LLM-powered conversational voice diary that delivers clinically grounded morning and evening sleep diary questions through proactive smart-speaker prompts, structured conversational intake, and adaptive follow-up dialogue. We evaluated the system in a four-week between-subjects field study with 30 university students, comparing it with a text-based mobile diary using matched diary items, reporting windows, and reminder intervals. Compared with the text-based diary, the conversational voice diary showed higher adherence and elicited more detailed contextual self-report about routines, stressors, environmental conditions, and other sleep-related factors. Participants also described the voice diary as easier to integrate into daily routines, despite longer perceived completion time. However, voice-based conversational intake produced lower completeness for some structured diary fields, revealing a trade-off between expressive richness and structured precision. These findings show both the promise and the challenge of using LLM-powered conversational voice assistants for longitudinal health self-report. 
\end{abstract}

\begin{teaserfigure}
  \includegraphics[width=\textwidth]{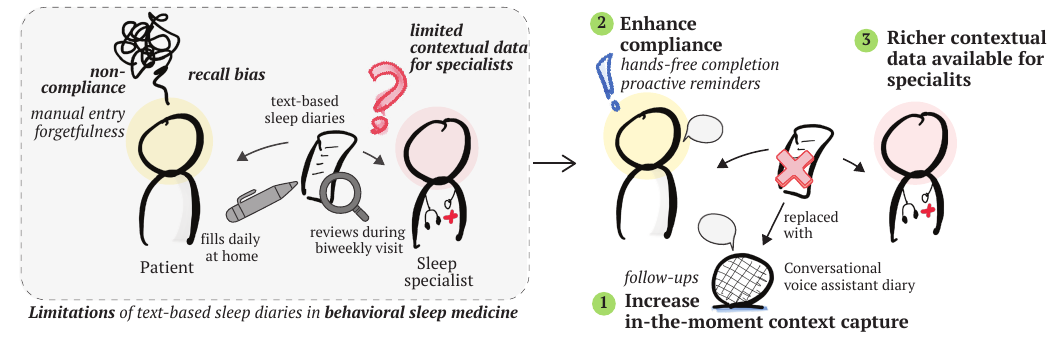}
  \caption{In this work, we propose a conversational sleep diary to address key limitations of traditional text-based sleep diaries in behavioral sleep medicine. We designed and developed an LLM-powered voice assistant for sleep diary intake. We compare our LLM-powered voice assistant diary with a standard text-based diary to showcase its efficacy and validity.}
  \label{fig:teaser}
\end{teaserfigure}


\maketitle

\section{Introduction}

Sleep is a fundamental component of physical, cognitive, and emotional well-being, and poor sleep is associated with elevated cardiovascular risk, cognitive impairment, depression, and diminished quality of life~\cite{morin2015insomnia,irwin2015sleep,lloyd2022lifes8}. Sleep also varies from night to night, shaped by daily stress, routines, schedules, and environmental conditions~\cite{harvey2008comparative,morin2015insomnia}. Understanding poor sleep and tracking changes over time therefore require accurate day-level assessment of both sleep patterns and the contextual factors surrounding them~\cite{carney2012consensus,morin2015insomnia}. To support this, sleep diaries are widely used in behavioral sleep medicine as a standard tool for prospectively capturing self-reported sleep. Standardized instruments such as the Consensus Sleep Diary (CSD) capture key variables, including bedtime, sleep onset latency, awakenings, wake time, and perceived sleep quality, through day-to-day reports completed close to sleep and wake events, providing ecologically valid subjective data that complement objective measures such as actigraphy~\cite{carney2012consensus,ancoli2003role}. Sleep diaries are particularly central to cognitive behavioral therapy for insomnia (CBT-I), the first-line recommended treatment for chronic insomnia~\cite{edinger2021behavioral,morin2015insomnia}, where they inform initial assessment, guide personalization of behavioral interventions such as sleep restriction and stimulus control, and track treatment response over time \cite{morin2015insomnia,edinger2021behavioral,carney2012consensus}.

Despite their clinical importance, sleep diaries have persistent challenges that can undermine their clinical utility. Adherence to daily completion is frequently compromised when patients miss entries, complete them outside the reporting window, or reconstruct nights retrospectively at clinic visits~\cite{stone2002patient,kristbergsdottir2023digital,tao2018eve}. Digital diaries introduced automated reminders and bounded entry windows~\cite{stone2003patient,hufford2002ecological}, but did not address the effort of manually entering information while groggy, fatigued, or rushed~\cite{datta2022sleep}. These lapses can be especially consequential among people with more severe symptoms or higher stress~\cite{bolger2003diary,aan2016compliance}, and missing or delayed entries can introduce recall bias that degrades metrics such as sleep efficiency, sleep onset latency, and wake after sleep onset~\cite{harvey2002,robbins2019,carney2012consensus}. Furthermore, structured forms often elicit limited contextual detail. Brief responses such as ``slept poorly'' or ``felt stressed'' reveal little about the specific stressors, routines, environmental disruptions, or coping behaviors behind a given night~\cite{ibanez2018survey,bolger2003diary}. Yet, such contextual information is important because behavioral sleep medicine depends not only on measuring sleep patterns, but also on interpreting the factors that shape night-to-night variability and identifying behaviors or routines that can be addressed in treatment~\cite{harvey2008comparative,morin2015insomnia}.


Recent advances in voice interfaces and large language models (LLMs) create an opportunity to address these limitations by rethinking sleep diary intake as a conversational, routine-embedded process rather than a static form-filling task. Voice-based reporting is hands-free and can fit naturally into morning and evening routines~\cite{porcheron2018voice,pradhan2020accessibility}, and speaking aloud may encourage elaboration that typing into a form does not~\cite{bickmore2005establishing,grimes2008toward,xiao2020}. A voice assistant deployed in the home can further prompt users within clinically appropriate reporting windows, supporting timely completion without requiring them to retrieve a phone, open an application, or remember the diary on their own~\cite{pielot2014understanding,zargham2022proactivity}. With LLM-powered dialogue, such systems can also interpret open-ended responses, ask follow-up questions when answers are vague, and support conversational repair~\cite{laranjo2018conversational,openai2023gpt4,sahijwani2022}, capturing both structured sleep variables and contextual information that helps explain sleep variation.

However, it remains unclear how LLM-powered voice diaries perform when used repeatedly in everyday home settings. Prior voice-based systems for sleep self-report and CBT-I have mainly examined feasibility in child populations~\cite{chen2024dozzz,aarts2022snoozy}, short-term non-LLM diary agents in adults~\cite{almzayyen2022vca}, or smart-speaker delivery of structured CBT-I content rather than diary intake itself~\cite{starling2024voicecbti}. Also, LLM-in-health research has largely focused on single-session tasks such as medical question answering rather than clinically grounded daily self-report~\cite{singhal2023medpalm,nori2023capabilities}. As a result, several questions remain about whether such a system can serve as a viable sleep diary intake tool. It is unclear whether a conversational voice diary can sustain daily use over weeks of field deployment, motivating \textit{RQ1: How does an LLM-powered conversational voice diary compare with a text-based diary in supporting day-to-day completion and usability over time?} Diary intake must also capture clinically meaningful structured variables while helping users explain the circumstances surrounding their sleep, motivating \textit{RQ2: How does conversational voice intake shape the quality and interpretability of self-reported sleep data?} Finally, a proactive voice diary operates in the home, where reporting is embedded in daily routines and shaped by shared spaces and varying comfort with speaking aloud, motivating \textit{RQ3: How do users experience a proactive conversational voice diary in their daily routines?}


To address these questions, we designed and deployed an LLM-powered conversational voice assistant (VA) on a smart speaker and evaluated it in a four-week between-subjects field study ($N=30$) against a text-based mobile diary that used the same items, reporting windows, and reminder schedule. The voice assistant delivered clinically grounded morning and evening sleep diary questions by combining proactive prompting, structured conversational intake, and adaptive follow-up dialogue. This paper makes the following contributions:

\begin{itemize}[leftmargin=*]
    \item \textbf{A clinically grounded LLM-powered conversational voice sleep diary system} that supports structured sleep diary intake through proactive prompting, conversational questions, and adaptive follow-up dialogue.
    \item \textbf{Longitudinal field evidence} from a four-week between-subjects study comparing conversational voice-based and text-based intake across adherence, usability, perceived completion time, structured completion, disclosure, and contextual richness.
    \item \textbf{Design implications} for conversational voice-based health diaries, including mechanisms for validating clinically important fields, supporting privacy-sensitive voice interaction in shared spaces, and adapting conversational pacing.
\end{itemize}

\section{Related Work}

Below, we review prior work on sleep self-report, voice-based health self-report, conversational elicitation, and LLM-powered health systems.

\subsection{Self-Report in Sleep Health and Behavioral Sleep Medicine}
Diary methods and ecological momentary assessment (EMA) are widely used in behavioral and clinical research to capture experiences close to when they occur, improving ecological validity and reducing retrospective recall bias compared with delayed questionnaires~\cite{bolger2003diary,shiffman2008ema,stone2002patient}. This temporal proximity is especially important for sleep, where recall can be shaped by fatigue, mood, and the salience of atypical nights~\cite{harvey2002}. Sleep diaries therefore ask people to report sleep and wake experiences soon after they happen, producing day-level subjective data about bedtime, sleep onset, awakenings, wake time, and perceived sleep quality~\cite{carney2012consensus}. These reports complement objective measures such as actigraphy and polysomnography~\cite{ancoli2003role,lauderdale2008self,robbins2019} and are widely used in behavioral sleep medicine for assessment, treatment planning, and tracking response to interventions such as cognitive behavioral therapy for insomnia ~\cite{carney2012consensus,morin2015insomnia,edinger2021behavioral}.


Prior sleep diary research has examined adherence, completion timing, missingness, delayed entry, and agreement or discrepancy between subjective diary estimates and objective measures, including work on digital diaries that introduced reminders, timestamps, and bounded entry windows~\cite{stone2003patient,hufford2002ecological,robbins2019,ancoli2003role,lauderdale2008self}. Despite these advances, sustained diary completion remains difficult; missed, delayed, or retrospectively reconstructed entries can introduce missingness and recall bias into longitudinal sleep data~\cite{stone2002patient,kristbergsdottir2023digital,tao2018eve}. Also, less attention has been given to how the diary intake format shapes the content of individual entries. Static, structured forms are effective for standardizing clinically important sleep variables, but they offer limited support for eliciting the circumstances surrounding a given night, including stressors, routines, environmental disruptions, coping behaviors, and other contextual factors that help explain night-to-night variability~\cite{ibanez2018survey,bolger2003diary,harvey2008comparative,morin2015insomnia}. Our work builds on this gap by examining whether conversational voice intake can support both timely completion and richer contextual sleep self-report while preserving the structured information needed for clinical interpretation.

\subsection{Voice-Based Health Self-Report}
Voice interfaces have been studied for health self-report across applications including symptom tracking, medication adherence, chronic disease management, and daily health logging~\cite{bickmore2006healthdialog,ermolina2021voice,almzayyen2022vca,millard2022alexa,maharjan2022}. Compared with manual or form-based entry, spoken interaction can reduce procedural effort by allowing users to respond without locating a diary, navigating fields, or writing out responses~\cite{almzayyen2022vca,millard2022alexa}. Smart speakers extend this interaction into domestic settings, where users can engage with an ambient device while moving through existing routines~\cite{sunshine2021smart,lopatovska2019talk,porcheron2018voice}. Home-based voice systems can also issue prompts within appropriate reporting windows, reducing reliance on users remembering to initiate entries on their own~\cite{pielot2014understanding,zargham2022proactivity}. Prior work further suggests that speaking aloud can support more elaborated descriptions of experiences, emotions, and reasons than short written responses typically capture~\cite{bickmore2005establishing,grimes2008toward}.

Sleep-related voice systems have begun to explore this design space. Prior work has examined voice-based sleep diaries across different populations and settings~\cite{chen2024dozzz,almzayyen2022vca}, as well as smart-speaker delivery of structured CBT-I content~\cite{starling2024voicecbti}. These studies establish the feasibility of voice interfaces for sleep-related self-report and intervention delivery, but they do not yet show how conversational voice-based sleep diary intake functions as a longitudinal, in-home self-report practice. We address this gap through a four-week field deployment of conversational voice-based sleep diary intake in participants' homes.

\subsection{Conversational Agents for Eliciting Self-Report}
Conversational agents can shape self-report not only by changing the input modality, but also by changing how information is elicited. Compared with static forms, conversational systems can acknowledge user responses, personalize prompts, ask clarifying questions, and respond when answers are vague or incomplete~\cite{car2020,provoost2017embodied,laranjo2018conversational}. Prior studies on embodied and relational conversational agents have shown that users may experience these systems as supportive and may disclose sensitive or personal information more readily than through traditional questionnaires~\cite{bickmore2005establishing,bickmore2010maintaining,bickmore2006healthdialog}. Research on conversational surveys similarly suggests that adaptive questioning can elicit more complete, informative, and descriptive responses than static web forms~\cite{sahijwani2022,skantze2021turn,xiao2020,jiang2023,kim2019comparing}. These findings are relevant to health self-report because the data people provide are shaped by how questions are asked, what they are prompted to recall, and how much space they are given to explain their experiences~\cite{choe2014understanding,epstein2015lived}.

Sleep diary intake requires users to translate subjective sleep experiences into standardized variables, such as sleep onset latency, number of awakenings, wake after sleep onset, and sleep quality~\cite{carney2012consensus}. However, clinically meaningful interpretation often depends on details beyond these variables, including stressors, routines, environmental disruptions, and other contextual factors~\cite{harvey2008comparative,morin2015insomnia}. Prior work has shown that structured diary formats are useful for standardizing sleep measurement, but they can provide limited space for capturing the circumstances surrounding a poor or unusual night~\cite{ibanez2018survey,bolger2003diary}. A conversational system could therefore ask targeted follow-up questions to clarify important diary responses while also eliciting contextual explanations that do not fit neatly into predefined form fields. Our work builds on this literature by examining how conversational voice-based reporting shapes sleep diary intake compared with static text-based reporting.

\subsection{LLM-Powered Conversational Agents in Health}

Large language models have expanded the design space for conversational health systems by enabling more flexible interpretation and generation of natural language~\cite{bommasani2021foundation,zhao2026survey,openai2023gpt4}. Compared with scripted systems, LLM-powered agents can better handle paraphrased responses, open-ended explanations, conversational repair, and unexpected user input without requiring every dialogue path to be authored in advance~\cite{laranjo2018conversational,sahijwani2022}. Within healthcare, LLMs have been studied for medical question answering, patient education, clinical summarization, and conversational support~\cite{singhal2023medpalm,wang2024applications,lee2023benefits}. For longitudinal self-report, these capabilities can support more flexible intake because users often provide approximate, narrative, or mixed responses that combine structured values with contextual explanations. However, much empirical work on LLMs in health has focused on single-session tasks, offline benchmarks, or curated clinical datasets~\cite{singhal2023medpalm,nori2023capabilities}. Existing studies provide limited evidence about how LLM-powered conversational systems function when used repeatedly in everyday settings for self-report. Our work examines an LLM-powered voice assistant for sleep diary intake, where the model supports conversational interaction over clinically grounded diary questions and asks adaptive follow-ups when responses are vague, incomplete, or contextually meaningful.

\section{Conversational Sleep Diary: System Design and Implementation}
\label{sec:system}
Motivated by the limitations of static sleep diaries and the potential of conversational voice-based reporting, we developed an LLM-powered conversational voice assistant for sleep diary intake. The system supports morning and evening diary entries through spoken interaction, proactive prompts within personalized reporting windows, and adaptive follow-up questions.

\subsection{Design Rationale}
\label{sec:design-rationale}

Sleep diary intake is a repeated, time-sensitive self-report task whose value depends on whether entries are completed close to sleep and wake events, whether clinically meaningful variables are captured consistently, and whether users can explain the circumstances surrounding a given night. We therefore designed the system around four considerations:

\textbf{Lowering the effort of repeated reporting.}
We chose voice as the primary interaction modality because text-based diary entry requires users to access a form, navigate fields, and type responses, which can make sleep diary completion feel like a separate task during low-energy moments. The assistant asks diary questions aloud and accepts spoken responses, allowing users to complete entries while still in bed, getting ready, or moving through morning and evening routines.

\textbf{Supporting time-proximal completion.}
To support entries close to the relevant sleep or wake event, the system proactively prompts users during personalized morning and evening reporting windows. Users can also initiate the diary themselves during the allowable reporting window. Once an entry is completed, the system stops prompting for that diary window to avoid duplicate entries.

\textbf{Preserving clinically grounded structure.}
Although the interaction is conversational, sleep diary intake still needs to collect structured information used in behavioral sleep medicine. We therefore retained a predefined set of morning and evening diary questions covering core sleep timing, perceived sleep quality, sleep aid use, naps, caffeine and alcohol intake, medication use, physical activity, stress, fatigue, sleep environment, and bedtime routines. The assistant phrases these questions in natural language rather than clinical terminology, while the underlying structure remains tied to clinically meaningful diary fields.

\textbf{Eliciting context for interpretation.}
Structured fields can show that a night was poor or disrupted, but they often provide limited insight into why. The assistant therefore asks adaptive follow-up questions when a response is vague, incomplete, or suggests that additional context would be useful. For example, if a user says they woke up ``a few times,'' the assistant can ask for an approximate number of awakenings; if a user rates their sleep quality as low, the assistant can ask what contributed to the poor sleep. These follow-ups clarify underspecified responses and elicit contextual details while preserving the structured diary flow.

\subsection{System Implementation}

Our implementation builds on the dual-API architecture introduced by Mahmood et al.~\cite{mahmood2025user}, integrating OpenAI's GPT-4o with the Alexa Skills Kit within an AWS Lambda function. When the skill is invoked, the Lambda handler forwards the user's utterance into our backend pipeline, where a middleman API coordinates communication and helps satisfy platform timing constraints before relaying the request to the backend~\cite{mahmood2025user}. The backend then calls GPT-4o for language understanding, response generation, dialogue management, and structured sleep diary status updates. We selected GPT-4o for its conversational capabilities, robust handling of paraphrases and follow-ups, and support for system-level prompts that constrain outputs. The backend and supporting APIs manage session state, timestamps, conversation history, and data logging, while structured diary data are stored in per-participant Google Sheets. Below, we describe the system's configurable parameters, state tracking mechanism, and functional modules.

\begin{figure*}[t]
  \includegraphics[width=\textwidth]{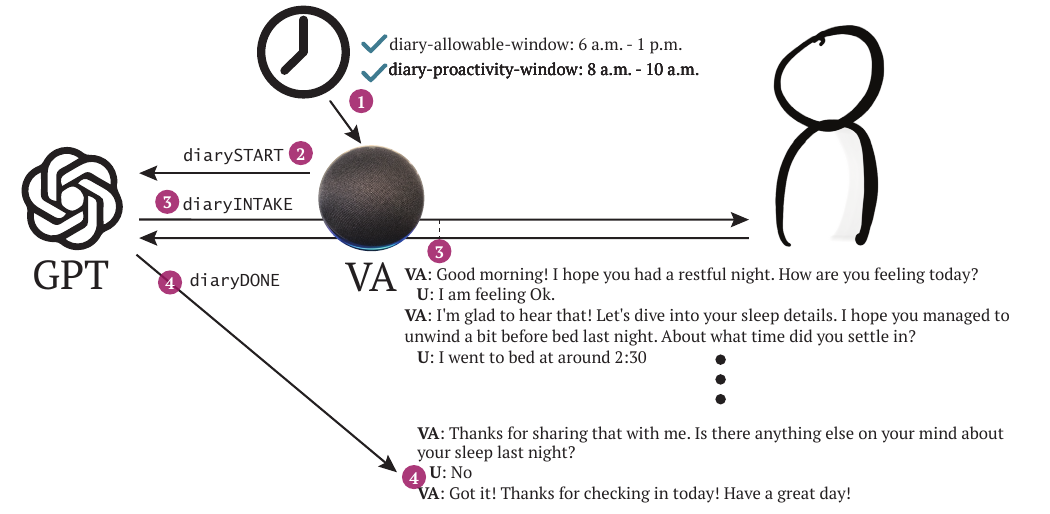}
  \caption{Interaction flow of the LLM-powered sleep diary system.}
  \Description{System interaction flow diagram.}
  \label{fig:interaction-flow}
\end{figure*}

\subsubsection{Configurable parameters}
The system uses two time-related parameters to determine when sleep diary intake (morning or evening) can occur and when proactive prompts are issued. These parameters are designed to accommodate individual differences in sleep schedules while adhering to clinical best practices for sleep diary completion.

\begin{itemize}[leftmargin=*]
  \item \texttt{diary-allowable-window}: The time window during which morning or evening diary entries can be completed and considered valid. This parameter is grounded in clinical practice and expert-consensus standards, which emphasize completing sleep diaries near sleep and wake events to ensure accuracy and temporal relevance \cite{carney2012consensus}. Entries completed outside this window are flagged as potentially subject to recall bias. In our deployment, we set a broad allowable window of 6:00 AM–1:00 PM for morning diaries and 7:00 PM–2:00 AM for evening diaries to accommodate the variable schedules of college students, who may have late-night study sessions or irregular wake times due to academic demands.

  \item \texttt{diary-proactivity-window}: A personalized time span (a two-hour window) during which the system may proactively prompt the user to complete the diary if it has not yet been submitted. This window is tuned per user based on their reported typical sleep and wake times, collected during the study setup session. Within this window, prompts are issued at a fixed reminder interval. In our deployment, the reminder interval was set to 15 minutes, balancing the need to remind users without being overly intrusive. For example, if a user typically wakes at 8:00 AM, their morning proactivity window might be set to 8:00–10:00 AM, with prompts at 8:00, 8:15, 8:30, and so on until the diary is completed or the window ends.
\end{itemize}

\subsubsection{State tracking}

The system maintains a single state variable, \texttt{sleepDiarySTATUS}, to track the progress of diary completion. This variable encodes both the diary type (morning or evening) and the interaction stage, enabling the system to resume interrupted sessions and prevent duplicate entries. Specifically, \texttt{sleepDiarySTATUS} can take one of three stages—\texttt{START}, \texttt{INTAKE}, or \texttt{DONE}—with a \texttt{morning} or \texttt{evening} prefix (e.g., \texttt{morningSTART}, \texttt{eveningINTAKE}, \texttt{morningDONE}).

\begin{enumerate}[leftmargin=*]
  \item \texttt{START}: The system is ready to begin the diary intake and is awaiting user readiness. This state is entered when the current time falls within the \texttt{diary-allowable-window} and no entry has yet been completed for that window.
  
  \item \texttt{INTAKE}: The diary intake is in progress. The system is actively asking questions and collecting responses. This state allows the system to track which questions have been answered and which remain, enabling session resumption if the user is interrupted.
  
  \item \texttt{DONE}: The diary intake has been completed for the current window. This state prevents further prompts during that window and signals that the data have been logged.
\end{enumerate}

State transitions are governed by both system logic and the interaction flow. When the skill is activated within the \texttt{diary-allowable-window} and no entry has yet been completed, the system sets \texttt{sleepDiarySTATUS} to the appropriate \texttt{START} state (e.g., \texttt{morningSTART} if invoked in the morning). Upon entering the \texttt{START} state, GPT-4o initiates the diary questions and the state transitions to \texttt{INTAKE}. Once all required questions are answered, the system sets the state to \texttt{DONE}, logs the entry with a completion timestamp, and prevents further prompts during that window. When the skill is invoked, \texttt{sleepDiarySTATUS} is used to determine whether to start a new diary intake, resume an interrupted session, or close the interaction if the diary has already been completed.

This state management design serves two purposes. First, it ensures that users can resume interrupted sessions without losing progress. For example, if a user starts the morning diary but is interrupted by a phone call, they can return to the diary within 15 minutes and continue from the last unanswered question. Second, it prevents duplicate entries, which could introduce noise into the data and confuse users. 

\subsubsection{Modules}

The system consists of two functional modules that orchestrate the diary intake process: the Activation Module, which determines when and how the diary is initiated, and the Interaction Module, which manages the conversational flow and data collection.

\textbf{Activation Module.} This module determines whether and how the diary intake is initiated based on the system parameters and \texttt{sleepDiarySTATUS}. The skill can be activated in two ways. In system-initiated (proactive) activation, within the \texttt{diary-proactivity-window}, Alexa proactively prompts the user at the configured reminder interval (every 15 minutes) with a brief tone and the message: ``Hi, it's time for your sleep diary intake. Say `I am ready' when you are.'' In user-initiated activation, users can say ``Alexa, sleep diary'' at any time. If the request occurs outside the \texttt{diary-allowable-window} or if the diary entry has already been completed, Alexa responds with ``It is not time for your sleep diary yet. Take care,'' and closes the session. If \texttt{sleepDiarySTATUS} is already \texttt{INTAKE}, the module resumes the conversation from the last question asked.
 
\textbf{Interaction Module.} Powered by GPT-4o, this module manages the conversational flow of the diary intake through two behaviors. First, \textit{structured diary questions}: a predefined set of clinically grounded sleep diary questions (see Appendix~\ref{app:sleep-diary-questions}) are asked in natural language based on the Consensus Sleep Diary~\cite{carney2012consensus}. Each morning, users report bed time, lights-out time, sleep onset latency, whether they went to bed as planned, number of awakenings, total wake time during the night (WASO), final awakening time, out-of-bed time, any early awakening deviations, sleep quality on a 1--10 scale, and any sleep aids used. Each evening, users report naps, alcohol and caffeine intake, medication use, physical activities, evening activities, pre-bed stress on a 1--10 scale, daytime fatigue on a 1--10 scale, sleep environment details, and bedtime routines. Questions are phrased conversationally (\eg ``About how long did it take you to fall asleep after you turned off the lights?'') rather than using clinical terminology. Second, \textit{dynamic follow-up questions}: GPT-4o generates follow-up questions to clarify vague responses, probe for missing details, or elicit additional contextual information. For example, if a user reports their sleep quality as ``3,'' the system might ask ``What made your sleep poor last night?'' These follow-ups are generated only when the initial response is insufficient for clinical interpretation or when additional context would be valuable. Once all required information is collected, the VA asks whether the user has anything else to add, then sets \texttt{sleepDiarySTATUS} to \texttt{DONE} and ends the session.

\section{User Study}


We conducted a four-week, between-subjects longitudinal study comparing two sleep diary intake methods: a text-based mobile diary (Text condition, $N=15$) and an LLM-powered voice assistant diary (VA condition, $N=15$). 

In the \textbf{Text condition}, participants used a Google Forms-based sleep diary accessible via a link sent through email. The diary presented structured questions in a scrollable form format, allowing participants to type free-text responses and select from drop-down menus for categorical variables. Entries were timestamped automatically upon submission. For consistency,  participants received proactive email reminders every 15 minutes within a personalized two-hour window (morning and evening) until they completed the diary. Participants could also access the diary link at any time within the allowable reporting window (6:00 AM--1:00 PM for morning diaries, 7:00 PM--2:00 AM for evening diaries) by clicking the link from any previous reminder email.

In the \textbf{VA condition}, participants used the Alexa-based sleep diary system described in Section ~\ref{sec:system}, which combined structured clinical questions with adaptive conversational follow-ups and proactive voice-initiated reminders. The proactive reminder window, reminder interval (15 minutes), and allowable reporting window were matched to the Text condition to ensure comparability.

Participants were assigned to conditions based on the order of device pick-up at the setup session: the first 15 participants to pick up their study materials were assigned to the VA condition, and the next 15 were assigned to the Text condition.


\subsection{Hypotheses}


Drawing on our system design goals and prior work on sleep diaries, voice-based health reporting, and conversational information elicitation, we formulated five hypotheses comparing the conversational VA diary with the text-based diary.

\textbf{Hypothesis 1 (Adherence).} The conversational VA diary will lead to higher adherence compared to text-based diaries. 
This hypothesis is informed by prior work showing that hands-free interaction and proactive reminders reduce interaction friction and support routine integration in daily contexts \cite{porcheron2018voice,pradhan2020accessibility,almzayyen2022vca,zargham2022proactivity,pielot2014understanding}. In contrast, traditional diary modalities require deliberate effort and are associated with higher rates of missing or delayed entries \cite{stone2002patient,kristbergsdottir2023digital}.

\textbf{Hypothesis 2 (Completion Time).} The conversational VA diary entries will require more time to complete than text-based entries. 
This hypothesis is motivated by the sequential and turn-based nature of spoken interaction, which introduces conversational overhead such as clarification, turn-taking, and repair \cite{moore2017conversational,wienrich2021,car2020}. While voice reduces physical effort, it may increase interaction duration compared to compact text forms.

\textbf{Hypothesis 3 (Perceived Usability).} The conversational VA diary will result in higher perceived usability than text-based diaries. 
This hypothesis is informed by prior work showing that conversational agents can improve user experience through natural language interaction, reduced cognitive load, and increased accessibility, particularly in low-energy states such as morning or bedtime routines \cite{bickmore2005establishing,pradhan2020accessibility,lewin2024novel}.

\textbf{Hypothesis 4 (Engagement and Disclosure).} The conversational VA diary will lead to higher user engagement and greater comfort in disclosing sleep-related experiences. 
This hypothesis builds on prior research showing that conversational agents can establish rapport, provide empathetic feedback, and encourage self-disclosure compared to static questionnaires \cite{bickmore2010maintaining,provoost2017embodied,xiao2020}. Although voice interaction may introduce privacy concerns in shared environments \cite{lau2018always,abdi2019scoping}, we expect the conversational nature of the system to support increased engagement overall.

\textbf{Hypothesis 5 (Contextual Richness).} The conversational VA diary entries will capture more sleep-related contextual factors than text-based entries. 
This hypothesis is informed by prior work demonstrating that conversational elicitation and adaptive follow-up questions can scaffold recall and increase the informativeness and specificity of user responses compared to static forms \cite{xiao2020,jiang2023,sahijwani2022,grimes2008toward}.

\subsection{Data Collection}
We collected the following data during the 4-week study:
\begin{enumerate}[leftmargin=*]
    \item \textit{Sleep diary entries.} 
    
    \textbf{VA condition:} In addition to Alexa usage logs and automated conversation logging in the backend, we employed an open-source recording device \cite{mahmood2024situated} to capture participants' interactions with the Echo Dot. An Arduino Nano 33 BLE Sense monitored the Echo's blue light ring, triggering audio capture to a microSD card; recording continued until 10 seconds after the light turned off, with extensions if Alexa was reactivated during that window. A red LED indicator illuminated only during active recording to ensure participant awareness. The functionality was confirmed via two two-week pilot deployments in real homes. We collected approximately 80 hours of audio data.
    
    \textbf{Text condition:} All diary entries were logged to a secure Google Forms backend with timestamps. Each entry included participant ID, submission timestamp, and responses to all structured questions and free-text fields. Email delivery logs confirmed successful reminder delivery.

    
    \item \textit{Post-study questionnaires.} We administered questionnaires at the end of study to assess user perceptions of system usability and effectiveness. 
    \item \textit{Semi-structured interviews.} We conducted semi-structured post-study interviews to capture user experience and perceptions of the diary system, routine integration, and long-term sustainability.
\end{enumerate}


\subsection{Data Analysis and Metrics}

For the VA condition, we transcribed, corrected and augmented the data collected through automated conversation logging in the backend by cross-referencing it with Alexa usage logs and interaction audio data collected from the recorder. We then automatically segmented the diary entries for each day using the tags ``start morning sleep diary'' and ``start evening sleep diary''. Each segment was timestamped and contained the complete user-VA conversation. For all 15 participants, their morning and evening segments were extracted in this format. A researcher manually reviewed the segments to clean missed entries and ensure the accuracy of segmentation.

For the Text condition, diary entries were directly extracted from the Google Forms backend. Each entry was already segmented by submission (one form submission per diary entry) and included all structured and free-text responses. 

We constructed the following metrics based on interaction data, and questionnaires.

\subsubsection{Adherence: completion rate}
Completion rates were calculated for the diaries. A diary was considered complete only when all questions had been presented and answered.

\subsubsection{Usability}

We evaluated the usability of the system through the System Usability Scale (SUS) score \cite{brooke1996sus} from the post-study questionnaire. 

\subsubsection{Efficiency: perceived time to complete} 
Perceived time to complete diaries was estimated through two participant reported questions: 1) Usually how long did it take for you to finish the morning sleep diary intake (in minutes) and 2) Usually how long did it take for you to finish the evening sleep diary intake (in minutes). 




\subsubsection{Diary entry depth}
We evaluated diary entry depth through a combination of quantitative and qualitative metrics:



\noindent \textbf{Quantitative metrics: information density}
We used linguistic measures to estimate the amount,
diversity, structural complexity, and idea density of information participants provided in each diary entry. For the VA condition, measures were calculated from participant speech only within each
diary segment. For the text-based condition, measures were calculated from participant-entered responses only. We used SciPy, spaCy, and NLTK, and applied a predefined exclusion list for the VA condition to remove interactional commands and device-directed phrases such as ``stop,''``thank you,'' ``continue,'' and ``Alexa'' e.t.c.

\begin{enumerate}[leftmargin=*]

\item \textbf{Response volume.} How much content participants
produced.
  \begin{itemize}
    \item \textit{Number of tokens:} Total count of tokens produced by the tokenizer. Numeric tokens (e.g., durations, times) are included as they are meaningful in the context.
    \item \textit{Words in noun phrases:} Total number of tokens that belong to noun phrases detected by the spaCy parser \texttt{noun\_chunks}.
  \end{itemize}
  
\item \textbf{Lexical diversity.} How varied and content-bearing the vocabulary was.
  \begin{itemize}

    \item \textit{Lexical density (\%):} Proportion of tokens that are content words (nouns, verbs, adjectives, adverbs). Spoken language tends to show lower lexical density than written text \cite{castello2008text}.

    \item \textit{Shannon entropy (bits):} Information-theoretic
    entropy of the token distribution, computed as
    \(H = -\sum_i p_i \log_{2} p_i\), after case normalization and
    exclusions. Higher values correspond to more varied or less
    repetitive vocabulary, used as a proxy for informativeness
    \cite{xiao2020}.
    \item \textit{Type--Token Ratio (TTR):} Ratio of unique tokens to total tokens.
    \item \textit{Moving-average TTR (MATTR):} Length-adjusted lexical diversity, computed as the average TTR across fixed-size sliding windows (window size = 50). Less sensitive to text length than raw TTR.
    \item \textit{Average word length:} Mean number of characters per token, used as a proxy for lexical sophistication.
  \end{itemize}

\item \textbf{Syntactic complexity.} How structurally complex the
language was.
  \begin{itemize}
    \item \textit{Mean dependency distance:} Average distance between syntactically dependent tokens across all sentences.
    \item \textit{Maximum tree depth:} Maximum depth of the dependency parse tree across all sentences.
    \item \textit{Cross entropy (bits):} Average negative
    log-likelihood (in bits per token) assigned by GPT-2 to the text, reflecting the unpredictability or surprisal of phrasing (lower values indicate greater predictability).
  \end{itemize}

\item \textbf{Idea density.} How many distinct ideas were expressed per unit of text.
  \begin{itemize}
    \item \textit{Proposition density:} Approximate number of
    proposition markers (tokens with subject or clausal dependency
    labels) per 10 words. Used as a heuristic measure of idea density.
    \item \textit{CPIDR-style idea-density score:} The Computerized Propositional Idea Density Rater (CPIDR) estimates propositional idea density from part-of-speech-tagged text~\cite{brown2008automatic}. Following this framing, we calculated an approximate CPIDR-style
    score as the number of finite verbs and clausal dependencies per 10 words.
  \end{itemize}

\end{enumerate}

\noindent \textbf{Qualitative metrics: human coding}
To assess the depth and clinical relevance of diary entries, we developed a coding scheme with four dimensions:

\begin{enumerate} [leftmargin=*]
    \item \textbf{Disclosure level (1--5):} Captures the depth and specificity of information disclosed in a diary entry, ranging from basic factual statements (Level 1) to unsolicited contextual or narrative detail (Level 5). This dimension is conceptually aligned with prior measures of specificity and descriptive richness \cite{xiao2020}. Detailed definitions and examples of each level are summarized in Table~\ref{tab:disclosure-levels}.

    \item \textbf{Completeness level (0--2):} Assesses whether each required sleep diary field contains sufficient information for clinical interpretation:  
    0 = missing or no usable information,  
    1 = vague or underspecified information (e.g., ``right away'' for sleep latency),  
    2 = sufficient detail for clinical use (e.g., ``5--10 minutes''), similar to relevance in \cite{xiao2020}.
    While disclosure level captures expressive depth, completeness directly evaluates the clinical validity of the reported information.

    \item \textbf{Engagement level:} A composite metric defined as Disclosure~$\times$~Completeness. This measure reflects both how much participants chose to say and whether they provided the necessary information. Higher scores indicate entries that are both expressive and clinically complete.

    \item \textbf{Sleep-related contextual factors:} Counts of spontaneously mentioned stressors, facilitators, coping behaviors, or environmental influences (e.g., noise, temperature, light). These contextual elements offer clinically relevant insights into day-to-day variability in sleep and are meaningful for behavioral sleep medicine assessment \cite{carney2012consensus,harvey2008comparative,morin2015insomnia}.

\end{enumerate}

\noindent \textbf{Inter-rater reliability}
To ensure coding reliability for the qualitative metrics, two researchers independently coded diary entries from three participants (10\% of data). 
Prior to independent coding, both coders reviewed a detailed codebook with representative examples to establish a shared understanding of the coding criteria. 

For ordinal measures (\textit{completeness} and \textit{disclosure})
inter-rater reliability was assessed using Intraclass Correlation Coefficients (ICC) with a single-measurement, consistency, two-way mixed-effects model (ICC(3,1)), yielding ICC(3,1) = .989 for \textit{completeness} and ICC(3,1) = .971 for \textit{disclosure} \cite{koo2016guideline}, indicating excellent intercoder reliability. For categorical sleep-related contextual factors, reliability was assessed using Cohen’s Kappa, yielding $\kappa = .946$ \cite{mchugh2012interrater}, reflecting almost perfect agreement.

Following the reliability assessment and resolution of disagreements, the two coders each completed coding for half of the remaining entries using the finalized codebook.

\begin{table}[ht]
\centering
\resizebox{\linewidth}{!}{%

\begin{tabular}{c p{6cm} p{6cm}}
\textbf{Disclosure Level} & \textbf{Definition} & \textbf{Example(s)} \\
\toprule
1 & \textbf{Basic factual.} Provides only direct, objective, or minimally descriptive responses without additional context. & 
``Yes''; ``woke up at 8 a.m. in the morning''; ``got out of bed right away'' \\
\hline
2 & \textbf{Contextual but non-specific.} Adds brief descriptive or contextual information, but remains vague and does not include detailed actions or clear causes. & 
``Yes I managed to get into the bed around 1:30 a.m. and it took me a while to fall into sleep''; 
``My mood was pretty good but I was very busy and a bit stressed''
 \\
\hline
3 & \textbf{Specific narrative.} Describes concrete experiences, actions, or feelings with situational detail, but without explicit causal explanation. & ``Yes my day went pretty well and I feel great''; ``I feel terrible I worked every waking minute of the day''; ``was going to sleep right away but then I called my I called my boyfriend and I stayed up for a bit talking to him''
 
 \\
\hline
4 & \textbf{Causal explanation.} Provides explicit reasons or causal relationships explaining sleep experiences. & ``Yes it was a late night due to a assignment it took me around 10 to 15 minutes to fall asleep so maybe around 1:15''; ``I think it was because I didn't get that much sleep 2 nights ago.''
\\
\hline
5 & \textbf{Additional or unsolicited information.} Introduces new, relevant information beyond what was directly prompted. & ``I woke up at 8 a.m. in the morning I plan to sleep during the afternoon for a quick nap''; ``I helped with a fundraiser for a club that I'm in which was interesting.''

\end{tabular}
}
\caption{Disclosure levels with definitions and illustrative excerpts from participant responses.}
\label{tab:disclosure-levels} 
\end{table}


\subsubsection{Interviews: perceptions}

We analyzed the transcribed post-study interviews using reflexive thematic analysis~\cite{braun2023doing} to contextualize participants' experiences with each diary modality. Two authors conducted iterative open coding and discussed recurring patterns across conditions; the resulting themes were used to organize the interview findings.


  \begin{figure*}[t!]
     \includegraphics[width=\textwidth]{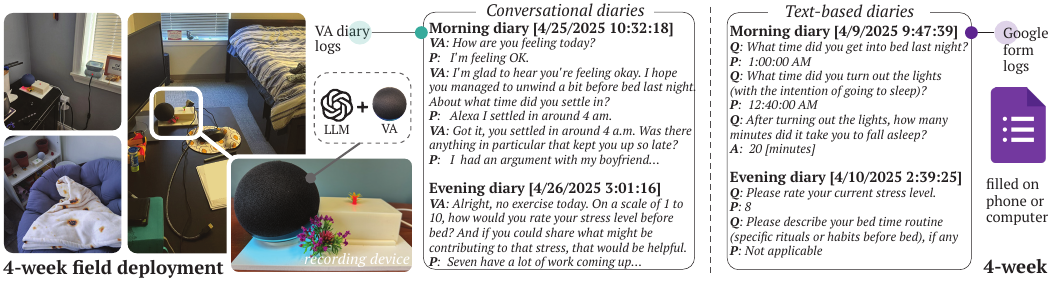}    
     \caption{4 week deployment and data collection.}
    \Description{....}
    \label{fig:3d-printing}
\end{figure*}

\subsection{Longitudinal Study: Procedure}

The study comprised a four-week field deployment that included a pre-study setup session for consent, baseline assessments, and device distribution, a mid-study check-in at the start of Week 3, and a final debrief session at the end of Week 4 involving a semi-structured exit interview.

\subsubsection{Pre-Study Setup Session}
Prior to the start of Week 1, participants attended an in-person setup session consisting of the following activities:

\begin{enumerate}[leftmargin=*]
    \item \textit{Introduction and consent.} The experimenter explained the study procedures, data collection methods, and privacy protections, and obtained written informed consent.
    
    \item \textit{Sleep-window profiling.} The experimenter documented participants' typical wake-up times (weekday and weekend), departure and return-home times, and usual wind-down period to configure personalized morning and evening proactive reminder windows (corresponding to the system parameters \texttt{diary-allowable-window} and \texttt{diary-proactivity-window} described in Section ~\ref{sec:system}), and then verified these details with participants.
    
    \item \textit{Study setup and training.} The experimenter provided a hands-on tutorial on how to complete sleep diary entries based on the assigned condition. In the VA condition, participants practiced using the Alexa sleep diary skill by completing a sample morning and evening entry. 

    \item \textit{Device setup (VA condition only).} Participants in the VA condition were shown how to set up the Echo Dot smart speaker and the audio recorder in their home. The experimenter demonstrated that the recorder captures audio only when the red LED is illuminated, removed the microSD card, and played back sample recordings to verify that participants understood what would be recorded during their interactions. Participants were asked to set up the device within the next two days and share a photograph of the setup location via email. An instruction manual and technical support contact details were provided.


\end{enumerate}


\subsubsection{Week 1--4: Sleep Diary Intake} 
Week 1 began once participants confirmed they had completed the device setup (VA condition) or received the first email reminder (Text condition). Starting on Day 1 of Week 1, participants were prompted to complete morning and evening sleep diary entries every day for four weeks. In both conditions, proactive reminders were sent every 15 minutes within the personalized two-hour reminder window until the diary was completed. Participants could also initiate diary completion at any time within the allowable reporting window.


\subsubsection{Week 4: Final Session and Exit Interview}

At the end of Week 4, participants attended a final in-person session to return study materials (VA condition: Echo Dot and audio recorder; Text condition: no physical materials). Participants first completed a comprehensive battery of post-study questionnaires, including the System Usability Scale (SUS) \cite{brooke1996sus}, NASA Task Load Index (NASA-TLX) for cognitive load \cite{hart1988development}, and custom Likert-scale assessments of fit to routine, repetitiveness, disclosure comfort, motivation, sleep self-awareness, and perceived completion time for morning and evening diaries. This was followed by a semi-structured exit interview (approximately 30--45 minutes) probing their overall experience, routine integration, adherence challenges, disclosure comfort, voice-assistant interaction nuances (VA condition), and suggestions for future enhancements. Interviews were audio-recorded with participant consent and transcribed for analysis.

\subsection{Participants}

We recruited 30 college students (undergraduate and graduate) using university mailing lists and flyers around the campus. The participants (17 female, 13 male) were aged between 20-32 ($M=23.56$, $SD=3.15$) and had a variety of educational background. Most participants self-reported native English fluency ($N=23$), followed by advanced professional fluency ($N=6$); one participant reported only basic fluency. Participants reported limited prior experience with voice assistants ($M = 2.83$, $SD = 1.39$) and with smart speaker-based voice assistants ($M = 2.27$, $SD = 1.41$), measured on a 5-point scale where 1 indicated no prior experience and 5 indicated extensive experience.


\section{Findings}

\subsection{Adherence}

A one-tailed t test assuming unequal variances indicated that adherence (completion rates) in the VA condition ($M = 92.00, SD = 7.82$) was significantly higher than in the Text condition ($M = 70.79, SD = 28.54$), $t(16.09) = 2.78, p = .007$.


\subsection{Usability}
Because the data were not normally distributed, a nonparametric test was conducted. A one-tailed Mann–Whitney U test (Wilcoxon rank-sum test) indicated that SUS scores were significantly higher in the VA condition ($M=77.00$, $SD=13.50$) than in the Text condition ($M=68.75$, $SD=12.55$), $Z = -1.73$, $p = .042$, $W = 170$.

\begin{table*}[htbp]
\centering
\caption{Mixed-model repeated measures ANOVA results. Predictors were condition (between-subjects), diary type (within-subjects), and day (within-subjects). $^{\ast}p \leq .05$, $^{\ast\ast}p \leq .01$, and $^{\ast\ast\ast}p \leq .001$.}
\label{tab:anova-results}

\resizebox{\textwidth}{!}{%
\begin{tabular}{l l l l | l l l l}
\textbf{Effect} & $F$ & $p$ & \boldmath$\eta_p^2$ &
\textbf{Effect} & $F$ & $p$ & \boldmath$\eta_p^2$ \\
\hline

\multicolumn{4}{l}{\textbf{Num Tokens}} &
\multicolumn{4}{l}{\textbf{MATTR}} \\
Condition   & $F(1, 28.01)=65.52$ & $<.001^{***}$ & $0.701$  &
Condition   & $F(1, 28.02)=91.51$ & $<.001^{***}$ & $0.766$ \\
& \multicolumn{3}{l}{\textit{text (M = 23.95, SD = 6.58) $<$ VA (M = 99.26, SD = 49.82)}} 
& \multicolumn{3}{l}{\textit{text (M = 0.55, SD = 0.14) $<$ VA (M = 0.73, SD = 0.08)}} & \\
Diary type  & $F(1, 1401.39)=45.99$ & $<.001^{***}$ & $0.032$  &
Diary type  & $F(1, 1402.15)=183.43$ & $<.001^{***}$ & $0.116$ \\
& \multicolumn{3}{l}{\textit{evening (M = 59.84, SD = 49.44) $<$ morning (M = 69.25, SD = 55.21)}} 
& \multicolumn{3}{l}{\textit{evening (M = 0.68, SD = 0.10) $>$ morning (M = 0.62, SD = 0.17)}} & \\
Day         & $F(33, 1401.57)=1.73$ & $.006^{**}$ & $0.039$ &
Day         & $F(33, 1402.64)=1.22$ & $.182$ & $0.028$ \\
\hline

\multicolumn{4}{l}{\textbf{Lexical Density (\%)}} &
\multicolumn{4}{l}{\textbf{Average Word Length}} \\
Condition   & $F(1, 28.02)=342.63$ & $<.001^{***}$ & $0.924$  &
Condition   & $F(1, 28.02)=86.68$  & $<.001^{***}$ & $0.756$ \\
& \multicolumn{3}{l}{\textit{text (M = 76.89, SD = 7.44) $>$ VA (M = 55.48, SD = 6.38))}} 
& \multicolumn{3}{l}{\textit{text (M = 2.83, SD = 0.80) $<$ VA (M = 3.75, SD = 0.30)}} & \\
Diary type  & $F(1, 1402.23)=248.05$ & $<.001^{***}$ & $0.150$ &
Diary type  & $F(1, 1402.16)=268.10$ & $<.001^{***}$ & $0.161$ \\
& \multicolumn{3}{l}{\textit{evening (M = 67.78, SD = 13.37) $>$ morning (M = 62.92, SD = 11.52)}} 
& \multicolumn{3}{l}{\textit{evening (M = 3.55, SD = 0.51) $>$ morning (M = 3.11, SD = 0.87)}} & \\
Day         & $F(33, 1402.74)=1.02$ & $.441$ & $0.023$ &
Day         & $F(33, 1402.65)=1.14$ & $.268$ & $0.026$ \\
\hline

\multicolumn{4}{l}{\textbf{Shannon Entropy (bits)}} &
\multicolumn{4}{l}{\textbf{Proposition Density}} \\
Condition   & $F(1, 28.02)=176.96$ & $<.001^{***}$ & $0.863$  &
Condition   & $F(1, 28.01)=148.86$ & $<.001^{***}$ & $0.842$ \\
& \multicolumn{3}{l}{\textit{text (M = 3.28, SD = 0.64) $<$ VA (M = 5.38, SD = 0.66)}} 
& \multicolumn{3}{l}{\textit{text (M = 1.80, SD = 1.66) $<$ VA (M = 4.31, SD = 0.86)}} & \\
Diary type  & $F(1, 1401.50)=0.00$ & $.975$ & $0.000$ &
Diary type  & $F(1, 1402.64)=15.02$ & $<.001^{***}$ & $0.011$ \\
& \multicolumn{3}{l}{} 
& \multicolumn{3}{l}{\textit{evening (M = 3.28, SD = 1.22) $>$ morning (M = 3.03, SD = 2.22)}} & \\
Day         & $F(33, 1401.73)=2.53$ & $<.001^{***}$ & $0.056$ &
Day         & $F(33, 1403.29)=1.20$ & $.199$ & $0.027$ \\
\hline

\multicolumn{4}{l}{\textbf{Words in NPs}} &
\multicolumn{4}{l}{\textbf{Mean Dependency Distance}} \\
Condition   & $F(1, 28.01)=73.61$ & $<.001^{***}$ & $0.724$  &
Condition   & $F(1, 28.01)=97.97$ & $<.001^{***}$ & $0.778$ \\
& \multicolumn{3}{l}{\textit{text (M = 4.02, SD = 5.14) $<$ VA (M = 36.11, SD = 20.31)}} 
& \multicolumn{3}{l}{\textit{text (M = 1.09, SD = 0.23) $<$ VA (M = 2.13, SD = 0.54)}} & \\
Diary type  & $F(1, 1401.44)=21.58$ & $<.001^{***}$ & $0.015$ &
Diary type  & $F(1, 1401.44)=40.24$ & $<.001^{***}$ & $0.028$ \\
& \multicolumn{3}{l}{\textit{evening (M = 22.68, SD = 20.88) $>$ morning (M = 19.98, SD = 23.27)}} 
& \multicolumn{3}{l}{\textit{evening (M = 1.70, SD = 0.63) $>$ morning (M = 1.60, SD = 0.70)}} & \\
Day         & $F(33, 1401.64)=1.86$ & $.002^{**}$ & $0.042$ &
Day         & $F(33, 1401.65)=1.96$ & $.001^{**}$ & $0.044$ \\
\hline

\multicolumn{4}{l}{\textbf{Cross Entropy (bits)}} &
\multicolumn{4}{l}{\textbf{Max Tree Depth}} \\
Condition   & $F(1, 28.01)=0.27$ & $.609$ & $0.010$ &
Condition   & $F(1, 28.02)=151.84$ & $<.001^{***}$ & $0.844$ \\
& \multicolumn{3}{l}{} 
& \multicolumn{3}{l}{\textit{text (M = 2.03, SD = 1.21) $<$ VA (M = 4.77, SD = 1.45)}} & \\
Diary type  & $F(1, 1402.67)=718.83$ & $<.001^{***}$ & $0.339$ &
Diary type  & $F(1, 1402.35)=67.87$ & $<.001^{***}$ & $0.046$ \\
& \multicolumn{3}{l}{\textit{evening (M = 5.83, SD = 0.96) $>$ morning (M = 4.73, SD = 0.70)}} 
& \multicolumn{3}{l}{\textit{evening (M = 3.76, SD = 1.78) $>$ morning (M = 3.25, SD = 2.01)}} & \\
Day         & $F(33, 1403.33)=0.36$ & $1.000$ & $0.008$ &
Day         & $F(33, 1402.90)=1.87$ & $.002^{**}$ & $0.042$ \\
\hline

\multicolumn{4}{l}{\textbf{TTR}} &
\multicolumn{4}{l}{\textbf{CPIDR-style idea-density score}} \\
Condition   & $F(1, 28.02)=30.92$ & $<.001^{***}$ & $0.525$ &
Condition   & $F(1, 28.01)=134.79$ & $<.001^{***}$ & $0.828$ \\
& \multicolumn{3}{l}{\textit{text (M = 0.55, SD = 0.14) $<$ VA (M = 0.66, SD = 0.09)}} 
& \multicolumn{3}{l}{\textit{text (M = 1.95, SD = 1.82) $<$ VA (M = 4.61, SD = 0.94)}} & \\
Diary type  & $F(1, 1402.10)=246.14$ & $<.001^{***}$ & $0.149$ &
Diary type  & $F(1, 1402.58)=15.39$ & $<.001^{***}$ & $0.011$ \\
& \multicolumn{3}{l}{\textit{evening (M = 0.65, SD = 0.10) $>$ morning (M = 0.57, SD = 0.14)}} 
& \multicolumn{3}{l}{\textit{evening (M = 3.52, SD = 1.30) $>$ morning (M = 3.24, SD = 2.40)}} & \\
Day         & $F(33, 1402.57)=0.76$ & $.832$ & $0.018$ &
Day         & $F(33, 1403.21)=1.91$ & $.212$ & $0.027$ \\
\hline

\multicolumn{4}{l}{\textbf{Completeness}} &
\multicolumn{4}{l}{\textbf{Disclosure}} \\
Condition & $F(1, 28.04)=74.01$ & $<.001^{***}$ & $0.725$ &
Condition & $F(1, 28.04)=55.95$ & $<.001^{***}$ & $0.666$ \\
& \multicolumn{3}{l}{VA (M = 1.65, SD = 0.22) < text (M = 1.92, SD = 0.09)} 
& \multicolumn{3}{l}{text (M = 0.97, SD = 0.05) < VA (M = 1.20, SD = 0.28)} & \\
Diary type & $F(1,1363.94)=1.21$ & $.272$ & $0.001$ &
Diary type & $F(1,1365.71)=0.03$ & $.853$ & $0.000$ \\
Day         & $F(67, 1363.10)=0.73$  & $.949$ & $0.035$  &
Day         & $F(67, 1364.00)=0.81$  & $.869$ & $0.038$ \\
\hline

\multicolumn{8}{l}{\textbf{Engagement Level}} \\
Condition   & $F(1, 28.00)=52.80$      & $<.001^{\ast\ast\ast}$ & $0.653$  & & & & \\
& \multicolumn{3}{l}{text (M = 1.93, SD = 0.09) < VA (M = 2.37, SD = 0.56)} \\
Diary type  & $F(1, 28.00)=9.39$       & $.005^{\ast\ast}$      & $0.251$  & & & & \\
& \multicolumn{3}{l}{\textit{evening (M = 2.21, SD = 0.29) $>$ morning (M = 2.09, SD = 0.29)}} \\
Day         & $F(67, 1364.06)=0.77$ & $.915$                 & $0.036$  & & & & \\
\hline

\end{tabular}
} 
\end{table*}

\subsection{Efficiency: Perceived time to complete}


A mixed-model repeated measures ANOVA was conducted for perceived completion time with \textit{condition} as a between-subjects factor (Text vs. VA) and \textit{diary type} as a within-subjects factor (morning vs. evening). Perceived completion time was collected retrospectively in the post-study questionnaire as two participant-reported estimates, one for morning diaries and one for evening diaries. We found a significant main effect of condition, $F(1, 26.8) = 18.60, p < .001$. Perceived completion time was significantly shorter in the Text condition ($M = 3.13,\; SD = 2.00$) than in the VA condition ($M = 6.06,\; SD = 2.54$). No significant main effect of diary type and no condition-by-diary-type interaction were observed.





\subsection{Diary entry depth}

\begin{figure*}[t]
  \includegraphics[width=\textwidth]{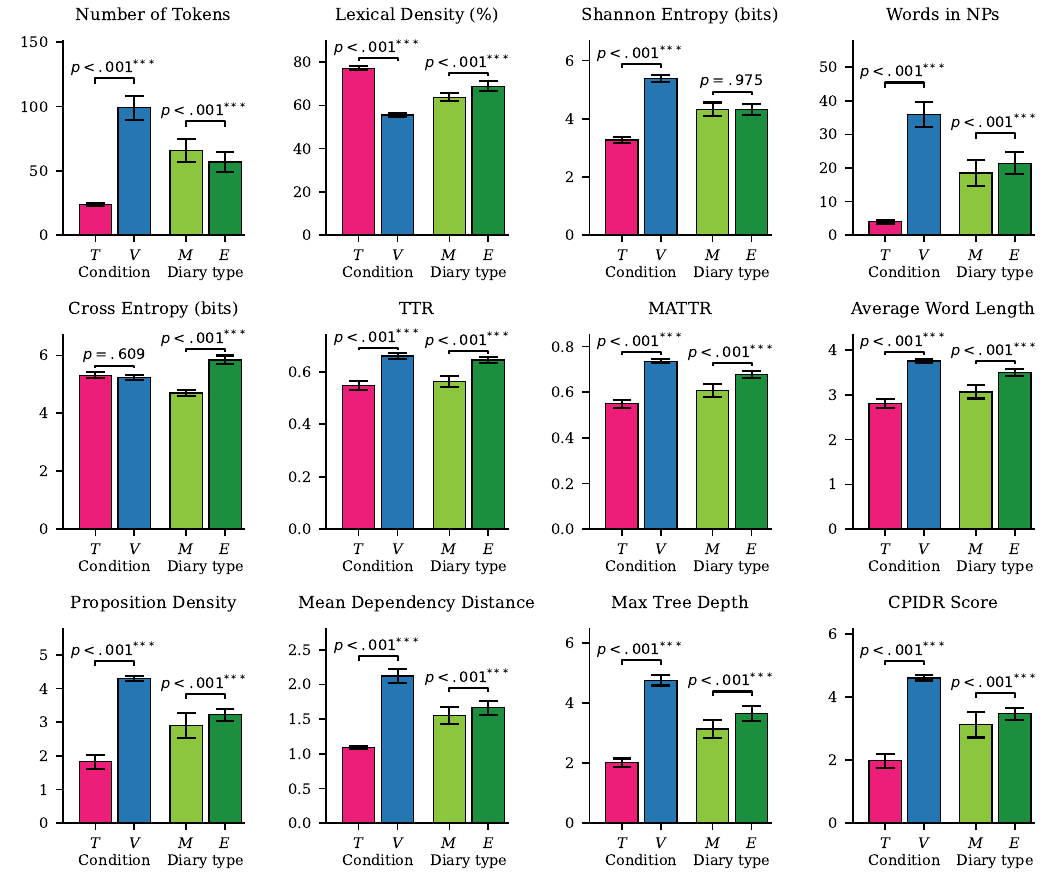}
  \caption{Diary entry depth: quantitative information-density metrics}
  \Description{}
  \label{fig:auto-metrics}
\end{figure*}


A series of mixed-model repeated measures ANOVAs were conducted for each dependent variable, with \textit{condition} as a between-subjects factor (Text vs. VA), and \textit{diary type} (morning vs. evening) and \textit{day} as within-subjects factors. No interaction terms were included. We fit additive models because our primary comparisons focused on average differences by condition, diary type, and day across diary-entry metrics, rather than on whether these effects varied across combinations of factors. Effect sizes are reported as partial eta squared ($\eta_p^2$) with 95\% confidence intervals. Following Cohen's guidelines, we considered $\eta_p^2=0.01$ a small effect, $\eta_p^2=0.06$ a medium effect, and $\eta_p^2=0.14$ a large effect~\cite{cohen1988statistical}. Results are presented in Table~\ref{tab:anova-results}.


\subsubsection{Quantitative metrics: information density}
\label{sec:quant-metrics}
\begin{itemize}
\item\textbf{Response volume was substantially higher in the VA condition.} 
Across both volume measures, participants produced more content in the VA condition than in the Text condition: number of tokens (VA $M = 99.26$ vs. 
Text $M = 23.95$, $p<.001$, $\eta_p^2=.701$) and words in noun phrases (VA $M = 36.11$ vs. Text $M = 4.02$, $p<.001$, $\eta_p^2=.724$). \textit{Diary type} had small but reliable effects on both measures ($p<.001$), and \textit{day} also contributed small effects ($p=.006$ for tokens; $p=.002$ for words in noun phrases).


\item\textbf{Lexical diversity was consistently higher in the VA condition.} 
Four of the five diversity measures showed large \textit{condition} effects favoring the VA condition: Shannon entropy ($p<.001$, $\eta_p^2=.863$), TTR ($p<.001$, $\eta_p^2=.525$), MATTR ($p<.001$, $\eta_p^2=.766$), and average word length ($p<.001$, $\eta_p^2=.756$). 
Lexical density showed the reverse pattern, with Text entries higher than VA ($M = 76.89\%$ vs. $M = 55.48\%$, $p<.001$, $\eta_p^2=.924$). This is consistent with text responses being shorter and more keyword-like, with fewer function words than spoken responses. \textit{Diary type} also had medium-to-large effects on most diversity measures (all $p<.001$ except Shannon entropy, $p=.975$); among these measures, evening entries showed slightly higher diversity than morning entries, likely because evening diaries asked about a broader range of daytime behaviors and contextual factors. \textit{Day} was significant only for Shannon entropy ($p<.001$, $\eta_p^2=.056$), suggesting day-to-day variation in lexical variety rather than a uniform trend across the diversity measures.

\item\textbf{Syntactic complexity was higher in the VA condition.} 
Participants in the VA condition produced more structurally complex language than those in the Text condition: mean dependency distance (VA $M = 2.13$ vs. Text $M = 1.09$, $p<.001$, $\eta_p^2=.778$) and maximum tree depth (VA $M = 4.77$ vs. Text $M = 2.03$, $p<.001$, $\eta_p^2=.844$). \textit{Diary type} also had small but reliable effects on both measures (both $p<.001$), with evening entries slightly more complex than morning entries, and \textit{day} contributed small effects ($p=.001$ for mean dependency distance; $p=.002$ for maximum tree depth). Cross entropy, which reflects 
phrasing predictability, did not differ by condition ($p=.609$) but showed a large diary-type effect ($p<.001$, $\eta_p^2=.339$), with evening entries less predictable than morning entries.

\item\textbf{Idea density was higher in the VA condition.} Both 
idea-density measures showed large condition effects: proposition density (VA $M = 4.31$ vs. Text $M = 1.80$, $p<.001$, $\eta_p^2=.842$) and CPIDR-style idea-density score (VA $M = 4.61$ vs. Text $M = 1.95$, $p<.001$, $\eta_p^2=.828$). \textit{Diary type} contributed small effects to both measures ($p<.001$), with evening entries slightly denser than morning entries. \textit{Day} was not significant for either measure.
\end{itemize}

\subsubsection{Qualitative metrics: human coding}
\label{sec:qual-metrics}




\begin{itemize}
\item \textbf{Completeness.} Completeness was significantly higher in the Text condition ($M = 1.92$, $SD = 0.09$) than in the VA condition ($M = 1.65$, $SD = 0.22$), $F(1, 28.04)=74.01$, $p<.001$, $\eta_p^2=.725$. This indicates that text entries more reliably contained the precise values needed for clinical interpretation. \textit{Diary type} was not significant, $F(1,1363.94)=1.21$, $p=.272$, and \textit{day} was not significant, $F(67,1363.10)=0.73$, $p=.949$.

\item \textbf{Disclosure.} Disclosure showed the opposite pattern, with VA entries ($M = 1.20$, $SD = 0.28$) scoring significantly higher than Text entries ($M = 0.97$, $SD = 0.05$), $F(1, 28.04)=55.95$, $p<.001$, $\eta_p^2=.666$. This indicates richer expressive depth and more contextual elaboration in spoken responses. \textit{Diary type} and \textit{day} were not significant, suggesting that the condition effect on disclosure was consistent across diary types and across the deployment.

\item \textbf{Engagement.} The composite engagement metric was significantly higher in the VA condition ($M = 2.37$, $SD = 0.56$) than in the Text condition ($M = 1.93$, $SD = 0.09$), $F(1, 28.00)=52.80$, $p<.001$, $\eta_p^2=.653$. Evening entries scored slightly higher than morning entries, $F(1,28.00)=9.39$, $p=.005$, $\eta_p^2=.251$; \textit{day} was not significant, $F(67,1364.06)=0.77$, $p=.915$, indicating stability over the deployment.
\end{itemize}

\begin{figure*}[!tp]
  \includegraphics[width=\textwidth]{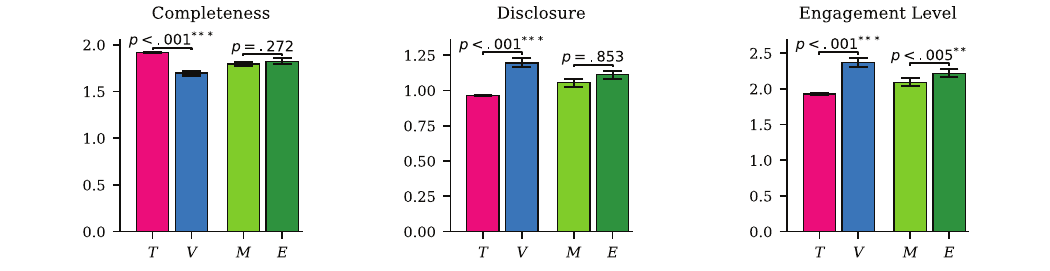}
  \caption{Diary entry depth: human-coded completeness, disclosure, and engagement}
  \Description{human-coded metrics.}
  \label{fig:human-coding}
\end{figure*}

\subsubsection{Sleep-related contextual factors}


We compared the number of coded sleep-related contextual factors mentioned in diary entries across conditions, including behavioral, physiological, psychological, work-related, and contextual factors. Overall, participants in the VA condition mentioned significantly more contextual factors per entry than those in the Text condition. The total number of tags was higher in the VA condition ($M = 14.67,\; SD = 14.76$) than in the Text condition ($M = 3.87,\; SD = 7.20$), $t(28) = 2.55$, $p = .017$, $d = 0.93$. Among the individual categories, only work-related factors showed a statistically significant difference, with substantially more work-related tags in the VA condition ($M = 4.20,\; SD = 4.72$) than in the Text condition ($M = 0.47,\; SD = 1.36$), $t(16.31) = 2.94$, $p = .009$, $d = 1.07$. Other categories showed numerical differences, but these were not statistically significant.

The tag distributions suggest that the two modalities elicited different kinds of contextual information. In the VA condition, frequently occurring tags were more likely to reflect explanations, coping strategies, and concrete causes of sleep disruption, including work, assignments, exams, tea, reading, white-noise machine use, thirst, and congestion. The examples in Table~\ref{tab:factor-examples} illustrate this pattern: participants often responded to follow-up questions by describing what happened, why it mattered, or what they did in response, such as staying up late because of work or drinking chamomile tea before bed. This made the VA entries more explanatory and easier to interpret.

By contrast, text entries were often much shorter and more elliptical. Participants frequently responded with brief phrases such as ``leave lab at 1:30 am, exhausted'' or ``First night after flight, a bit jet lagged,'' without explaining whether these factors helped or harmed sleep, how strongly they affected the night, or how they related to other behaviors. As a result, although a tag could be identified, the entry often lacked enough explanation to determine whether the factor functioned as a barrier, a facilitator, or simply a background condition. In addition, because participants often used similarly short wording across days, the text entries showed substantial repetition, especially for recurring environmental or routine-related descriptions. Together, these findings suggest that the VA did not just elicit more contextual factors overall, but more interpretable and varied ones, whereas the text diary more often produced short, repetitive mentions that were harder to contextualize.





\begin{table*}[t]
\centering
\footnotesize
\setlength{\tabcolsep}{4pt}
\renewcommand{\arraystretch}{1.12}
\caption{Illustrative examples of sleep-related contextual factors from the VA and Text conditions. The table is organized into two sections while preserving the same schema across both conditions: \textit{Tag Category}, \textit{Tag Name}, \textit{Example Question}, and \textit{Example User Answer}.}
\label{tab:factor-examples}
\resizebox{\linewidth}{!}{%

\begin{tabularx}{\textwidth}{p{1.8cm} p{3.2cm} p{4.8cm} Y}
\toprule
\textbf{Tag Category} & \textbf{Tag Name} & \textbf{Example Question} & \textbf{Example User Answer} \\
\midrule

\multicolumn{4}{l}{\textbf{VA Condition}} \\
\midrule
behavioral & COPING\_STRATEGIES\_TEA & Yes, the air quality can definitely impact your sleep. Thanks for mentioning that. Did you use any sleep aids last night, like medication or herbal teas? & I drank a chamomile tea before bed \\
physiological & BARRIER\_THIRST & That's pretty quick! Did you wake up at all during the night, or was it a smooth sleep through until morning? & Yes I was feeling thirsty and I woke up maybe around 3 or 4 a.m. but I don't remember the time \\
psychological & BARRIER\_DISCOMFORT & Oh, please continue. You were mentioning discomfort. Was it related to temperature, your mattress, or something else? & the discomfort was related to my mattress and congestion that I had \\
work & BARRIER\_WORK & I'm glad to hear that you're feeling alright. Let's dive into last night a bit. I hope you managed to unwind a bit before bed last night. About what time did you settle in? & Yes it took a longer time actually so I slept around 2 a.m. I was doing some work \\
contextual & FACTOR\_KAYAK & Got it, no medications today. Did you engage in any exercise or physical activity? & Yes, I went outdoors kayaking. \\

\midrule
\multicolumn{4}{l}{\textbf{Text Condition}} \\
\midrule
behavioral & FACTOR\_NONAP & Please describe your sleep environment (room temperature, light exposure, and noise levels) & Didn't take a nap \\
physiological & FACTOR\_ALCOHOL & Provide any important details about your sleep last night. & Drank a can of beer the previous night at a party, so had some headaches which led to trouble sleeping (Forgot to mention in previous night's diary) \\
psychological & FACTOR\_TIRED & Provide any important details about your sleep last night. & i was tired and sad. had headache. just wanted to sleep \\
work & FACTOR\_LAB & Provide any important details about your sleep last night. & leave lab at 1:30 am, exhausted \\
contextual & FACTOR\_JET & Provide any important details about your sleep last night. & First night after flight, a bit jet lagged \\
\bottomrule
\end{tabularx}
}
\end{table*}


\subsection{User perceptions: interview findings}

We conducted reflexive thematic analysis~\cite{braun2023doing} on the transcribed post-study interview data to examine how participants experienced each modality. Two authors conducted iterative open coding of the transcripts and organized recurring codes into seven dimensions of user experience: (1) integration into daily routine, (2) speed and pacing, (3) privacy and environmental sensitivity, (4) self-awareness and behavioral change, (5) disclosure comfort, (6) repetition, and (7) feedback and long-term value. Participants are identified by modality and sequential number (\eg VA-P02 for the second participant in the voice condition and TXT-P11 for the eleventh participant in the Text condition).

\subsubsection{Perceptions of integration into daily routine}
\label{sec:routine}
The two modalities integrated into participants' daily routines in distinct ways. A commonly noted benefit of the voice diary was the ability to complete entries while engaging in other activities. VA-P09 explained, \myc{``Because it was speech I could like multitask\ldots\ I could like do something in the kitchen while talking.''} Several participants described the voice diary as fitting naturally into existing routines with little deliberate effort. VA-P10 said, \myc{``Sometimes I would just like turn it on and do it while getting ready\ldots\ I just didn't really notice that I was just like doing it.''} VA-P08 similarly said, \myc{``Without having to even think about it, it just like happened naturally. So it just fitted really well.''} For VA-P10, the diary also became part of a bedtime transition, replacing phone use before sleep: \myc{``I would just go straight to sleep after finishing it\ldots\ wouldn't like check my phone or anything.''}

By contrast, participants in the Text condition more often described the diary as a standalone task requiring focused attention rather than something that blended into their existing routines. TXT-P13 said, \myc{``It felt like I was just inputting like similar things.''} One participant explicitly contrasted this with the imagined convenience of voice interaction. TXT-P11 said, \myc{``I feel like talking about the sleep would have been like easier to fit into a schedule cause I could have done it while doing something else.''} Routine integration was further complicated by the mobile interface: completing the diary on a phone was described as more cumbersome than on a laptop because of small form elements and the need to zoom in. TXT-P05 explained, \myc{``On the days that I was on my phone it was a little bit harder to fill out the form so that's when it started becoming like more like in the way of things.''} Other text participants similarly noted that the mobile form could be cumbersome when completing entries on a phone.

\subsubsection{Perceptions of speed and pacing}
\label{sec:pacing}
Despite its benefits for routine integration, the voice diary was often perceived as relatively slow because of its sequential turn-taking structure. VA-P02 described wanting to move through the diary more quickly but finding the one-question-at-a-time format frustrating: \myc{``I sometimes feel like I want to rush through the sleep diary but the fact that I have to like answer one question at a time and wait for it to respond\ldots\ was frustrating for me.''} VA-P02 estimated each session at approximately ten minutes. Other participants in the voice condition reported sessions of roughly five to eight minutes, and some perceived evening sessions as longer because of follow-up questions.

Participants in the Text condition generally reported faster completion times, often describing entries as taking about one or two minutes. TXT-P03 said, \myc{``It took me like a minute every time to complete.''} This efficiency came partly from the ability to see all questions at once and type responses without waiting for sequential prompts. However, this speed was accompanied by less conversational involvement and fewer opportunities for adaptive elaboration. Thus, the interviews align with the perceived-time results: text was faster, while VA provided a more interactive but slower reporting experience.

\subsubsection{Perceptions of privacy and environmental sensitivity}
\label{sec:privacy}
Voice interaction required speaking aloud, which introduced environmental and social considerations that were largely absent from text-based entry. Several participants described modifying their behavior to avoid disturbing others. VA-P04 said, \myc{``Because I sleep late I wasn't trying to talk super loud because I didn't want to disturb my roommates. So sometimes it would be like it just couldn't catch what I said.''} VA-P07 described whispering at night for similar reasons, which reduced the system's recognition accuracy. VA-P04 also recounted social situations where the device's notifications disrupted voice calls with friends: \myc{``Whenever it rings, like all my friends would make fun of me like, `Your Alexa is ringing in the background.'\,''}

Morning prompts were particularly intrusive when participants' sleep schedules shifted. VA-P03 said, \myc{``It was like early in the morning and I slept like at 3:00 and I didn't like want to wake up like 8:00 and started to tell, you know, `Hey, this is your diary. Let's start.'\,''} VA-P04 further raised concerns about unintended recording, noting that the device's indicator light sometimes remained active after the diary session ended: \myc{``I was just a little concerned that it recorded like parts of conversations that I didn't really want it to record.''}

Text-based entry did not raise comparable concerns about environmental visibility or social disruption. However, participants in the Text condition noted that email-based reminders created a different kind of intrusion, cluttering inboxes and occasionally being filtered into spam folders.

\subsubsection{Perceptions of self-awareness and behavioral change}
\label{sec:self-awareness}

Across both conditions, participants described increased awareness of their sleep-related habits, although this awareness emerged differently across modalities. Participants in the voice condition often described a form of anticipatory self-monitoring, in which knowing the diary would ask about specific topics led them to attend to those behaviors throughout the day. VA-P04 described noticing a caffeinated drink in the moment because she expected the diary to ask about it later: \myc{``I was drinking a matcha and I was like the sleep diary is gonna ask me about this. I'm gonna have to remember that I drink a caffeinated drink.''} VA-P07 noted that the exercise question prompted her to reconsider what counted as physical activity: \myc{``I might just do something active but wouldn't have considered it like a workout. But then when the sleep diary would be like, `Did you exercise today?' then like, `Oh yeah I guess.'\,''} VA-P08 similarly described the diary as helping surface patterns she might otherwise have overlooked.

Several VA participants also described speaking experiences aloud as reflective in itself. VA-P07 noted that \myc{``having to say it out loud in the morning and at night every day''} made sleep variability more salient, while VA-P10 described talking through stress as helping him realize that \myc{``things aren't like actually that bad.''}

Participants in the Text condition more often described awareness as emerging through the act of recording and noticing their own patterns. TXT-P05 identified a recurring relationship between caffeine intake and difficulty falling asleep: \myc{``On days where I drank caffeine it was just harder to fall asleep\ldots\ I was able to see that consistent pattern.''} TXT-P12 described being confronted with persistently poor sleep that she had previously normalized: \myc{``Every single day answering the question about my sleep quality and almost always marking it as low made me confront that.''}

For some participants, increased awareness was accompanied by behavioral changes. VA-P06 reported reducing screen time before bed, VA-P08 described adjusting room temperature and blinds to reduce nighttime awakenings, and TXT-P08 shifted toward an earlier bedtime. Several participants also framed the diary as a source of accountability. TXT-P03, for example, noted feeling \myc{``embarrassed saying I was sleeping past an unhealthy bedtime,''} which motivated more consistent sleep timing. At the same time, increased awareness was not always experienced positively. TXT-P15 described a negative feedback loop in which calculating sleep duration every morning heightened distress without leading to meaningful change: \myc{``I was sleeping less by just because I was calculating every morning how much I slept. But I was not able to change the routine so it made me feel bad about the way I sleep.''} These accounts suggest that increased awareness could be experienced as helpful or burdensome depending on whether participants felt able to act on what they noticed.

\subsubsection{Perceptions of disclosure comfort}
\label{sec:disclosure-comfort}
The extent to which participants disclosed personal information varied across conditions and was shaped by the system's responsiveness, the perceived audience, and the effort involved in elaborating. In the voice condition, empathetic system responses sometimes appeared to encourage deeper sharing over time. VA-P08 noted that feeling heard and validated made her more open: \myc{``Just like almost feeling heard and validated was helpful\ldots\ I think I was more open to like sharing.''} VA-P13 described the interaction as resembling a conversation with a friend: \myc{``I felt like it was I was talking to a friend.''}

This pattern was not consistent across all participants, however. VA-P12 reported becoming less willing to elaborate over time because she did not feel that the system was meaningfully using what she shared: \myc{``I didn't see a point in elaborating\ldots\ what anyone is getting from me being super descriptive.''}

Participants in the Text condition generally described more limited disclosure. Quantitative fields were easy to complete, but open-ended prompts often elicited short or minimal responses. TXT-P14 reported entering \texttt{N/A} in every \texttt{additional information} field because the prompt felt too vague: \myc{``The question is kind of a little bit vague, so I don't know what to put on it.''} TXT-P08 similarly explained that articulating the reasons behind her stress \myc{``would have been too much thinking.''}

Across both conditions, awareness that researchers would review responses emerged as a notable barrier to disclosure. VA-P04 noted uncertainty about who might listen to her recordings and whether she might be judged: \myc{``I wasn't sure who exactly would be listening to it\ldots\ I don't know if some people would judge me.''} TXT-P10 similarly described discomfort knowing that researchers would read the data. VA-P09 captured a distinction that also appeared in other interviews: \myc{``I was comfortable talking but I didn't feel comfortable sharing personal details.''} Together, these accounts indicate that disclosure comfort was shaped not only by modality but also by participants' perceptions of who would ultimately access their responses.

\subsubsection{Perceptions of repetition}
\label{sec:repetition}
Repetition was the most frequently reported concern across both modalities, though it manifested differently. In the voice condition, repetition was tied to the conversational flow: participants had to verbally respond to the same sequence of questions daily, including items that were consistently inapplicable. VA-P03 described the frustration of being asked about coffee and alcohol every day despite never consuming either: \myc{``I don't drink coffee, never, I don't drink alcohol. So but repetitively like asked every day, day and night.''} Several participants suggested that the system should adapt over time. VA-P10 said, \myc{``If somebody says no to medication for like two weeks maybe stop asking it every day and then like change it to weekly.''} VA-P12 expressed a desire for the system to develop persistent memory: \myc{``Having some sort of like memory that persists over time would have been really useful. Like I think referring back to previous conversations would have made it feel more like I was talking to a person.''}

In the Text condition, repetition was more closely tied to structural design issues. Participants who did not consume caffeine, alcohol, or medication were still required to select \texttt{no} and then type \texttt{NA} in the follow-up field each day. TXT-P10 said, \myc{``For all those fields every day I have to select like no and then typing NA which like over time felt very like repetitive and boring.''} TXT-P15 noted that the form lacked a button for not-applicable responses, requiring manual text entry each time. Several participants also reported that questions about sleep environment and bedtime routines generated identical answers daily because these aspects of their lives did not change.

Despite the frustration, some participants in both groups acknowledged that the predictability of the questions had a positive side: it allowed them to prepare answers throughout the day. VA-P05 called this a \myc{``double-edged sword,''} noting that she knew what to expect but grew tired of repeating the same answers. VA-P08 described actively thinking about what to report during the day because she anticipated the questions, which in turn heightened her awareness of sleep-related behaviors.

\subsubsection{Perceptions of feedback and long-term value}
\label{sec:feedback}
In both conditions, participants wanted clearer evidence that their repeated diary entries were useful, either through summaries, visualizations, or future clinical interpretation. The absence of any data visualization, summary, or insight was a recurring concern. VA-P01 said, \myc{``I'd like to see which were the days I had less sleep\ldots\ I think that's one of the reasons why Spotify Wrapped is really used.''} VA-P10 described wanting a companion application that would show trends over time, similar to fitness tracking platforms, and VA-P08 wished the system could provide weekly summaries or indicate whether her sleep patterns were improving.

Participants in the Text condition expressed similar desires. TXT-P14 said, \myc{``I'm actually interested how I'm sensitive to caffeine and stuff\ldots\ I'd like to see if that was one of the visualizations.''} TXT-P03 wanted a dashboard showing sleep duration alongside exercise and stress levels, with color-coding to highlight patterns, while TXT-P11 characterized the core limitation as the system being \myc{``input only''} with no analytical return.

Participants in both groups indicated that long-term use would be more sustainable if repeated self-report produced some visible value, such as personal summaries or clinically useful records. VA-P12 stated that she would not continue using the system in its current form because she felt the information she provided was not being utilized. Two text participants compared the diary to period-tracking applications and suggested that clinical integration, where the data could inform healthcare decisions, would provide a more compelling reason to continue. 

\section{Discussion}

We evaluated an LLM-powered conversational voice diary for sleep diary intake against a text-based mobile diary in a four-week field deployment. The findings show that conversational voice interaction changed not only how participants completed diary entries, but also what kinds of sleep-related information they shared. The VA condition was associated with higher adherence, longer perceived completion time, richer disclosure, and more contextual sleep-related information. At the same time, structured completeness was lower in the VA condition, revealing a trade-off between expressive flexibility and the precision of clinically important diary fields. Below, we discuss this trade-off, the role of conversational voice interaction in routine-embedded self-report, privacy and disclosure concerns in shared homes, and implications for future conversational self-report tools.

\subsection{From Structured to Conversational Self-Report}
Shifting from a static text-based diary to an adaptive conversational one substantially enriched the depth and contextual detail of what participants reported, while also surfacing a design trade-off between expressive flexibility and the precision of structured clinical fields.

\subsubsection{Capturing richer clinically relevant context}

The conversational diary elicited substantially more detailed and contextually rich self-report than the text-based diary (supporting H5). Entries in the VA condition contained more tokens, greater lexical diversity, and higher syntactic complexity across nearly all information density metrics (Section~\ref{sec:quant-metrics}, Table~\ref{tab:anova-results}). Human coding further revealed higher \textit{disclosure} and \textit{engagement} (Section~\ref{sec:qual-metrics}), with participants more frequently articulating explanations and contextual details alongside the required sleep variables (supporting H4). These findings are consistent with prior work on conversational surveys showing that dialog-based interaction supports more flexible and expressive reporting~\cite{xiao2020,laranjo2018conversational,sahijwani2022}, and extend those findings to the domain of longitudinal clinical self-report, where standard structured forms, while effective for capturing core metrics ~\cite{carney2012consensus}, constrain how people describe the situational factors that shape night-to-night sleep variability~\cite{ibanez2018survey}.

The interview data suggest that the adaptive structure of the conversational system played an important role. Participants in the voice condition explicitly described the back-and-forth exchange---follow-up questions, acknowledgments, and probes for vague responses---as encouraging them to elaborate on context they would not have typed into a form (Section~\ref{sec:disclosure-comfort}). This kind of turn-by-turn interaction can scaffold recall and encourage elaboration that a static text field does not afford. The resulting narratives, such as descriptions of stressors, environmental disruptions, coping behaviors, and emotional states, are precisely the contextual cues that help clinicians understand why sleep varied on a given night and identify behavioral targets for intervention~\cite{harvey2008comparative,morin2015insomnia}.

\subsubsection{Trade-off between completeness and disclosure}

Conversational flexibility also changed how structured information was expressed. Although the system was designed to preserve clinical structure by retaining a predefined set of clinically grounded morning and evening diary questions (Section~\ref{sec:design-rationale}), \textit{completeness} was lower in the VA condition than in the Text condition (Section~\ref{sec:qual-metrics}). Required clinical fields were more often conveyed in qualitative or approximate terms (e.g., ``a few times'' rather than a numeric count), even when the underlying question was unchanged. This pattern reflects a known tension between natural expression and precise value capture in dialog-based health reporting~\cite{car2020,wienrich2021}, where speech encourages elaboration but also accommodates vagueness in ways that a typed numeric field does not. The interview data further help explain this contrast. Participants in the VA condition often described the interaction as a back-and-forth exchange about their sleep, in which the system's follow-up questions and acknowledgments encouraged them to elaborate on context and stressors (Sections~\ref{sec:routine} and \ref{sec:disclosure-comfort}), whereas participants in the Text condition described diary completion as a focused logging task centered on filling in fields efficiently and correctly, with several explicitly noting that they minimized open-ended responses to save effort (Section~\ref{sec:disclosure-comfort}). The text format made structured fields easy and salient, while the conversational format made elaboration easy and salient.

Rather than a limitation of conversational intake, this trade-off points to a need for an additional verification layer on top of the structured questions. The current system was designed to ask clinically grounded questions but did not actively detect or repair underspecified responses, which allowed approximate spoken answers to pass through the dialogue. Future conversational diaries can address this by adding lightweight clarification, normalization, or end-of-entry confirmation strategies that detect missing or vague values for clinically critical fields and prompt the user to refine them~\cite{sahijwani2022,skantze2021turn}. 

\textit{Design implication.} Conversational diaries should pair adaptive elaboration with a targeted verification layer for fields with strict clinical interpretation requirements such as sleep onset latency, awakening counts, and wake time, so that expressive disclosure does not come at the cost of structured precision.

\subsection{What Conversational Voice Interactions Offer for Health Self-Reporting}
The conversational voice-based modality also reshaped how the diary fit into participants' daily lives.

\subsubsection{Integrating into existing routines} 

Participants using the conversational diary showed higher completion rates (supporting H1) and higher SUS scores (supporting H3) than those using the text-based diary, despite also reporting longer perceived completion times (supporting H2). This pattern suggests that adherence and usability are shaped not only by task duration but by how well an interaction fits into existing daily routines~\cite{bentley2018understanding}. The interview data reinforce this interpretation. Participants in the voice condition frequently described completing the diary while cooking, getting dressed, or winding down for bed, and several characterized it as a seamless part of their routine rather than a discrete task (Section~\ref{sec:routine}). For one participant, the diary became a bedtime transition that replaced phone use before sleep. By contrast, the text-based diary was more often described as a standalone task requiring focused attention, with mobile interface friction further increasing effort (Sections~\ref{sec:routine} and \ref{sec:pacing}). One participant in the Text condition explicitly imagined that voice interaction would have been easier to fit into a schedule because it could be done while multitasking. This pattern is consistent with prior work showing that hands-free voice interaction integrates more naturally into domestic routines than typed entry~\cite{sunshine2021smart,porcheron2018voice}, and suggests that the multitask-compatible nature of voice may be particularly well suited to twice-daily self-report tasks where users are often fatigued or cognitively occupied.

\subsubsection{Enabling reflection to foster awareness}

Both modalities increased participants' awareness of their sleep-related habits (Section~\ref{sec:self-awareness}), but the conversational voice diary appeared to support awareness through verbal reflection rather than only through pattern review. Participants in the voice condition described saying things aloud as making sleep variability more salient and as helping reframe perceived stress. This was largely absent from text-based reporting, where awareness emerged more analytically as participants reviewed their own entries and noticed patterns over time, such as a recurring caffeine--sleep relationship or persistently poor sleep quality that had previously been normalized. 
This distinction connects to prior work suggesting that verbalization scaffolds reflection in ways that writing does not~\cite{bickmore2005establishing,grimes2008toward}, and positions conversational voice diaries not only as data collection tools but also as lightweight reflective practices that may align with behavioral sleep medicine goals~\cite{choe2014understanding,epstein2015lived,morin2015insomnia}. However, heightened awareness was not uniformly positive.  One participant described a negative feedback loop in which daily tracking amplified distress without enabling change (Section~\ref{sec:self-awareness}), and many participants in both conditions wanted summaries, visualizations, or other forms of feedback that would translate awareness into action (Section~\ref{sec:feedback}). 

\textit{Design implication.} Conversational health diaries should provide visible value from repeated self-report without overwhelming users or turning diary intake into automated advice. Lightweight summaries or optional visualizations may help users understand the purpose of repeated tracking while reducing the risk that awareness becomes distressing.

\subsubsection{Surfacing privacy concerns in shared spaces} 
 

The voice modality and proactive prompting introduced environmental and social sensitivities largely absent from text-based entry (Section~\ref{sec:privacy}). Speech recognition errors were a recurring frustration, especially when participants lowered their voices to avoid disturbing roommates, which made the system less likely to recognize their responses accurately. Morning prompts were experienced as intrusive when sleep schedules shifted, social disruption from device notifications affected ongoing voice calls with friends, and concerns about unintended recording arose when the device's indicator light remained active longer than expected. These frictions are specific to spoken interaction in shared homes and have been documented in prior work on smart speakers and voice systems~\cite{lau2018always,abdi2019scoping,wienrich2021}. The persistence of these frictions across the four-week deployment suggests they are not simply early-onboarding issues but ongoing constraints on how voice-based health self-report can fit into shared living environments. 

\textit{Design implication.} Conversational voice diaries should expose user-facing controls over the social footprint of interaction---configurable prompt windows, snooze and deferral options, discreet prompt modalities, clear recording indicators, and easy fallback to text or whisper-tolerant input for sensitive contexts---so that the value of voice does not depend on having a private space available at every reporting moment.



\subsection{Disclosure Comfort and Perceived Data Security}

Beyond the social and environmental privacy frictions discussed above, participants' disclosure was also shaped by their beliefs about who would ultimately access their reported data. Among participants in the voice condition, empathetic system responses appeared to encourage deeper sharing for some, while others became less willing to elaborate over time when they felt their input was being recorded, retained, or not meaningfully used (Section~\ref{sec:disclosure-comfort}). Participants in the Text condition generally provided more limited open-ended responses, with several describing the prompts as vague or the act of typing as too effortful to support elaboration. The most frequently cited barrier to disclosure across both conditions, however, was not the interaction modality itself, but the awareness that researchers would review the responses. Participants using the voice diary expressed concern about who might listen to their recordings, while those using the text diary described discomfort knowing that others would read what they wrote.

This suggests that perceived audience may be a stronger determinant of disclosure depth than modality alone, an observation consistent with prior work on privacy perceptions in voice-based and digital health systems~\cite{lau2018always,malkin2019privacy}. Conversational interaction may encourage openness, but that openness remains shaped by users' beliefs about where their data goes and who ultimately encounters it. While the conversational structure often supported richer reporting, disclosure was still constrained by audience perceptions, especially when voice recordings felt more personal or socially legible than typed text. \textit{Design implication.} Authentic disclosure in conversational health systems depends not only on rapport or usability, but also on transparency and control over the data lifecycle. Supporting disclosure may therefore require clearer communication about who can access reported data, more user control over recording and storage, and audience-appropriate disclosure options that help users feel safe both technically and socially.

\subsection{Limitations and Future Work}

We acknowledge several limitations of this work. Our sample consisted of 30 university students who were not clinically diagnosed with insomnia or other sleep disorders. While participants exhibited variable sleep quality, their reporting behaviors, privacy sensitivities, and engagement patterns may not generalize to clinical populations with greater symptom severity or treatment motivation. Future work should evaluate conversational sleep diaries within clinical workflows and with broader demographic representation, including older adults and individuals with comorbid conditions for whom sleep tracking is particularly relevant.

Our four-week deployment, while sufficient for establishing stable group-level differences in adherence and reporting behavior, precludes conclusions about sustained engagement beyond this timeframe. Novelty effects may have contributed to the higher adherence and disclosure observed in the conversational condition, and whether these benefits persist over months of daily use remains an open empirical question~\cite{lazar2015critical}. Because participants were compensated and self-selected into the study, observed effect sizes may also be larger than would be expected in naturalistic deployment.

Our evaluation focused on participant adherence, perceived usability, diary-entry content, and user experience. We did not assess downstream clinical utility, including whether the richer contextual information captured by the conversational diary improves clinical interpretation, treatment planning, or patient--clinician communication. Future work should involve sleep medicine clinicians to examine how conversationally collected diary data should be summarized, reviewed, and integrated into behavioral sleep medicine workflows. Future iterations should also evaluate mechanisms for improving structured capture, such as missing-field detection, numeric confirmation prompts, normalization of approximate values, and end-of-entry value summaries.

Finally, we did not systematically examine LLM-specific risks such as hallucinated follow-ups, biased response phrasing, or inappropriate emotional responses in sensitive health contexts~\cite{wornow2023shaky,thirunavukarasu2023large}. Our prompt-based approach with GPT-4o was sufficient for generating design insights, but developing safeguards including output validation, human-in-the-loop oversight, and transparent communication of system limitations is necessary before broader clinical deployment~\cite{chancellor2019taxonomy,bickmore2018patient}.





\section{Conclusion}

We designed, deployed, and evaluated an LLM-powered conversational voice assistant for sleep diary intake, comparing it against a standard text-based diary in a four-week between-subjects field study with 30 participants. The conversational voice diary supported higher adherence and perceived usability and elicited richer contextual self-report across quantitative information density metrics and human-coded disclosure and engagement. Although completeness for structured clinical fields was lower in the voice condition, this reflects a need for an additional verification layer rather than a limitation of conversational intake itself. These findings show how adaptive conversational voice interaction can transform sleep diary intake from a transactional logging task into a reflective, routine-embedded practice that surfaces the situational context structured forms typically fail to capture.

\begin{acks} 
This work was supported by  Malone Center for Engineering in Healthcare. 
\end{acks}

\section*{CRediT Author Statement}
\textbf{Amama Mahmood}: Conceptualization, Methodology, Software, Validation, Formal analysis, Investigation, Data curation, Writing -- Original draft, Writing -- Review \& editing, Visualization, Project administration.
\\
\textbf{Bokyung Kim}: Conceptualization, Methodology, Software, Validation, Formal analysis, Investigation, Data curation, Writing -- Original draft, Writing -- Review \& editing.
\\
\textbf{Honghao Zhao}: Formal analysis, Data curation, Writing -- Original draft, Writing -- Review \& editing.
\\
\textbf{Luis F. Buenaver}: Conceptualization, Methodology, Writing -- Review \& editing.
\\
\textbf{Molly E. Atwood}: Conceptualization, Methodology, Writing -- Review \& editing.
\\
\textbf{Michael T. Smith}: Conceptualization, Methodology, Writing -- Review \& editing.
\\
\textbf{Chien-Ming Huang}: Conceptualization, Methodology, Resources, Writing -- Original draft, Writing -- Review \& editing, Visualization, Supervision, Funding acquisition.

\section*{Declaration of generative AI and AI-assisted technologies in the writing process}

During the preparation of this work the authors used ChatGPT in order to cut down repetitions and improve readability and language. After using this tool/service, the authors reviewed and edited the content as needed and take full responsibility for the content of the publication.



\newpage
\bibliographystyle{ACM-Reference-Format}
\bibliography{references}

\newpage
\input{appendix.tex}

\end{document}

%% file: appendix.tex
\appendix

\section{Sleep diary questions}
\label{app:sleep-diary-questions}
\textbf{Morning Sleep Diary}

\begin{enumerate}
    \item Time the user physically got into bed.
    \item Time the user tried to fall asleep (Lights-out time).
    \item How long it took them to fall asleep.
    \item Whether they went to bed as planned.
    \item Number of awakenings during the night.
    \item If the user indicates awakenings, total awake time (time spent awake across all awakenings).
    \item Time the user woke up for the day (Final Awakening).
    \item Time the user got out of bed to start the day
    \item Whether they woke earlier than planned, and if so, by how much time.
    \item Sleep quality rating on a 1–10 scale (1 = very poor, 10 = very good).
    \item Whether sleep aids were used, including types and timings.
\end{enumerate}

\textbf{Evening Sleep Diary}

\begin{enumerate}
    \item Any naps, including times and durations.
    \item Alcohol intake (include type, amount, and time).
    \item Caffeinated drinks intake (include type, amount, and times).
    \item Over-the-counter or prescription medication intake (include type and time).
    \item Exercise or physical activity (include type, duration, and time).
    \item Evening activities (screen time, reading, or relaxation techniques).
    \item Stress level before bed on a 1–5 scale (1 = very low, 5 = very high), including any major stressors.
    \item Fatigue level during the day on a 1–10 scale (1 = no fatigue, 10 = extreme fatigue).
    \item Sleep environment (room temperature, light exposure, and noise levels).
    \item Bedtime routine, if any (specific rituals or habits before bed).
\end{enumerate}